\documentclass[12pt, onecolumn, draftcls]{IEEEtran}
\usepackage{cite}%
\usepackage{bm}
\usepackage{geometry}
\usepackage{parskip}
\usepackage{psfrag}
\usepackage{subfigure}
\usepackage{url}
\usepackage{stfloats}
\usepackage{amsmath}%
\usepackage{array}
\usepackage{amssymb}%
\usepackage{float}
\usepackage{graphicx}
\usepackage{epstopdf}
\usepackage{bm}
\usepackage{color}
\newtheorem{thm}{Theorem}

\newtheorem{prop}{Proposition}

\DeclareMathAlphabet{\eurm}{U}{eur}{m}{n}
\DeclareMathAlphabet{\mathbsf}{OT1}{cmss}{bx}{n}
\DeclareMathAlphabet{\mathssf}{OT1}{cmss}{m}{sl}
\DeclareMathAlphabet{\mathcsf}{OT1}{cmss}{sbc}{n}

\newcommand{\randomvalue}[1]{\eurm{\uppercase{#1}}}

\DeclareSymbolFont{bsfletters}{OT1}{cmss}{bx}{n}  
\DeclareSymbolFont{ssfletters}{OT1}{cmss}{m}{n}
\DeclareMathSymbol{\bsfGamma}{0}{bsfletters}{'000}
\DeclareMathSymbol{\ssfGamma}{0}{ssfletters}{'000}
\DeclareMathSymbol{\bsfDelta}{0}{bsfletters}{'001}
\DeclareMathSymbol{\ssfDelta}{0}{ssfletters}{'001}
\DeclareMathSymbol{\bsfTheta}{0}{bsfletters}{'002}
\DeclareMathSymbol{\ssfTheta}{0}{ssfletters}{'002}
\DeclareMathSymbol{\bsfLambda}{0}{bsfletters}{'003}
\DeclareMathSymbol{\ssfLambda}{0}{ssfletters}{'003}
\DeclareMathSymbol{\bsfXi}{0}{bsfletters}{'004}
\DeclareMathSymbol{\ssfXi}{0}{ssfletters}{'004}
\DeclareMathSymbol{\bsfPi}{0}{bsfletters}{'005}
\DeclareMathSymbol{\ssfPi}{0}{ssfletters}{'005}
\DeclareMathSymbol{\bsfSigma}{0}{bsfletters}{'006}
\DeclareMathSymbol{\ssfSigma}{0}{ssfletters}{'006}
\DeclareMathSymbol{\bsfUpsilon}{0}{bsfletters}{'007}
\DeclareMathSymbol{\ssfUpsilon}{0}{ssfletters}{'007}
\DeclareMathSymbol{\bsfPhi}{0}{bsfletters}{'010}
\DeclareMathSymbol{\ssfPhi}{0}{ssfletters}{'010}
\DeclareMathSymbol{\bsfPsi}{0}{bsfletters}{'011}
\DeclareMathSymbol{\ssfPsi}{0}{ssfletters}{'011}
\DeclareMathSymbol{\bsfOmega}{0}{bsfletters}{'012}
\DeclareMathSymbol{\ssfOmega}{0}{ssfletters}{'012}

\newcommand{\rva}{{\randomvalue{a}}}	

\newcommand{\rvd}{{\randomvalue{d}}}	

\newcommand{\rvs}{{\randomvalue{s}}}	

\newcommand{\rvv}{{\randomvalue{v}}}	

\newcommand{\rvw}{{\randomvalue{w}}}	

\newcommand{\rvx}{{\randomvalue{x}}}	

\newcommand{\rvy}{{\randomvalue{y}}}	

\newcommand{\rvz}{{\randomvalue{z}}}	

\begin{document}

\title{Generalized Nearest Neighbor Decoding}
\author{Yizhu Wang, Wenyi Zhang
\thanks{Y. Wang and W. Zhang are with Department of Electronic Engineering and Information Science, University of Science and Technology of China. (Email: wenyizha@ustc.edu.cn)}
}

\maketitle

\begin{abstract}
It is well known that for Gaussian channels, a nearest neighbor decoding rule, which seeks the minimum Euclidean distance between a codeword and the received channel output vector, is the maximum likelihood solution and hence capacity-achieving. Nearest neighbor decoding remains a convenient and yet mismatched solution for general channels, and the key message of this paper is that the performance of nearest neighbor decoding can be improved by generalizing its decoding metric to incorporate channel state dependent output processing and codeword scaling. Using generalized mutual information, which is a lower bound to the mismatched capacity under independent and identically distributed codebook ensemble, as the performance measure, this paper establishes the optimal generalized nearest neighbor decoding rule, under Gaussian channel input. Several {restricted forms of the} generalized nearest neighbor decoding rule are also derived and compared with existing solutions. The results are illustrated through several case studies for fading channels with imperfect receiver channel state information and for channels with quantization effects.
\end{abstract}

\section{Introduction}
\label{sec:intro}

\subsection{Background and Motivation}
\label{subsec:motivation}

Information transmission over Gaussian channels has achieved tremendous success in the practice of digital communication. As a notable fact, the maximum likelihood and hence capacity-achieving decoder can be expressed as a nearest neighbor rule {\cite{lapidoth96:it}}, which seeks, for a given received channel output vector, the codeword that has the smallest Euclidean distance to it. Such a nearest neighbor decoding rule (NNDR) is simple and neat, providing a geometric interpretation for decoding. For channel $\rvy = \rvx + \rvz$ with length-$N$ codewords $x^N(m) = (x_1(m), \ldots, x_N(m))$, $m \in \mathcal{M} = \{1, \ldots, \lceil e^{NR}\rceil\}$, where the noise $\rvz$ is memoryless Gaussian and independent of $\rvx$, the NNDR is of the form {\cite{lapidoth96:it}}
\begin{eqnarray}
    \widehat{m} = \mathrm{arg} \min_{m \in \mathcal{M}} \sum_{n = 1}^N \left| y_n - x_n(m) \right|^2.
\end{eqnarray}
The NNDR can be extended to fading channels. For channel $\rvy = \rvs \rvx + \rvz$, where $\rvs$, the fading coefficient, is assumed to be perfectly known to the receiver, the NNDR is {\cite{lapidoth96:it}}
\begin{eqnarray}
    \widehat{m} = \mathrm{arg} \min_{m \in \mathcal{M}} \sum_{n = 1}^N \left| y_n - s_n x_n(m) \right|^2.
\end{eqnarray}

With the rapid evolution of wireless communication systems, as carrier frequency increases, bandwidth widens, and number of antennas increases, two critical issues become dominant. First, it is challenging for the receiver to gain the full knowledge of the fading process, and hence the channel state information (CSI) is generally imperfect. Second, it is costly and inefficient to realize perfectly linear transceivers, and nonideal transceiver distortion, as a consequence, is generally nonnegligible {\cite{bjornson14:it}}.

In the presence of such issues, the NNDR loses its optimality. Nevertheless, attributed to its simplicity and robustness, the NNDR has still been widely applied for channels beyond Gaussian, as a mismatched decoding rule.

For a memoryless channel without state, $p(y|x)$, $x, y \in \mathbb{C}$, when the NNDR is
\begin{eqnarray}
    \widehat{m} = \mathrm{arg} \min_{m \in \mathcal{M}} \sum_{n = 1}^N \left| y_n - \alpha x_n(m)\right|^2,
\end{eqnarray}
with scaling coefficient $\alpha = \mathbf{E}\left[\rvx^\ast \rvy\right]/\mathbf{E}\left[|\rvx|^2\right]$,\footnote{Throughout the paper, for a complex-valued vector, we use superscript $\ast$ to denote its conjugate transpose, and $|\cdot|^2$ to denote its norm.} and when the input $\rvx$ obeys independent and identically distributed (i.i.d.) circularly symmetric complex Gaussian distribution with mean zero and variance $P$, $\mathcal{CN}(0, P)$, an achievable information rate called the generalized mutual information (GMI) has been established in \cite[Appendix C]{zhang12com} as
\begin{eqnarray}
    I_\mathrm{GMI} &=& \log \frac{1}{1 - \Delta},\\
    \Delta &=& \frac{\left|\mathbf{E}\left[\rvx^\ast \rvy\right]\right|^2}{P \mathbf{E}\left[|\rvy|^2\right]}.
\end{eqnarray}

Furthermore, by allowing the receiver to process the channel output $\rvy$ in a symbol-by-symbol fashion before feeding it into the decoder, it has been shown in \cite{zhang16ita} \cite{zhang19jsac} that the GMI can be improved to
\begin{eqnarray}
    I_\mathrm{GMI, MMSE} &=& \log \frac{1}{1 - \Delta_{\mathrm{MMSE}}},\\
    \Delta_{\mathrm{MMSE}} &=& \frac{\mathrm{var} \mathbf{E}[\rvx|\rvy] }{P}.
\end{eqnarray}
This is achieved by modifying the NNDR into
\begin{eqnarray}
    \widehat{m} = \mathrm{arg} \min_{m \in \mathcal{M}} \sum_{n = 1}^N \left| \mathbf{E}[\rvx|y_n] - \alpha x_n(m) \right|^2,
\end{eqnarray}
with scaling coefficient $\alpha = \mathrm{var} \mathbf{E}[\rvx|\rvy]/P$; that is, we introduce an output processing function as the minimum mean-squared error (MMSE) estimate of the channel input upon observing the channel output.

{If we further allow the scaling coefficient to depend upon the channel ouput, how much can we improve the performance of NNDR? Besides, for channels with state, how to incorporate the CSI into the decoding process? These are what we treat in the present work.}

\subsection{Generalized Nearest Neighbor Decoding Rule (GNNDR)}
\label{subsec:gnnd}

\begin{figure}
    \centering
    \includegraphics[width=4.5in]{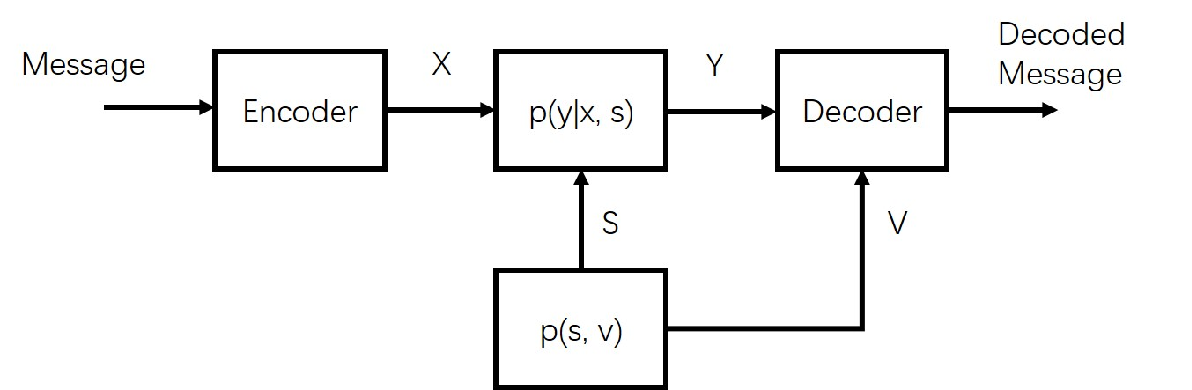}
    \caption{Illustration of system model.}
    \label{fig:model}
\end{figure}

We consider the general channel model illustrated in Figure \ref{fig:model}, in which the channel has a memoryless state $\rvs$ which governs the channel transition law $p(y|x, s)$, and also emits a CSI $\rvv$ available at the decoder. The special case of $\rvv = \rvs$ corresponds to perfect receiver CSI, and allowing $(\rvs, \rvv)$ to obey a general joint probability distribution $p(s, v)$ enables us to investigate imperfect receiver CSI. See Section \ref{sec:model} for further discussion.

Throughout the paper, we investigate the following generalized NNDR (GNNDR),
\begin{eqnarray}
\label{eqn:general-nndr}
    \widehat{m} = \mathrm{arg} \min_{m \in \mathcal{M}} \sum_{n = 1}^N \left|g(y_n, v_n) - f(y_n, v_n) x_n(m)\right|^2.
\end{eqnarray}
The mappings $g$ and $f$ are called the processing function and the scaling function, respectively. Their choices impact the resulting achievable information rate of the GNNDR.

In Section \ref{subsec:general-gmi}, we introduce GMI as the performance measure, and present the general expression of the GMI for the GNNDR (\ref{eqn:general-nndr}), under i.i.d. Gaussian codebook ensemble. Subsequently, in Section \ref{subsec:optimal-gnnd}, we derive the optimal pair of the processing function and the scaling function, which together maximize the GMI. What plays the key role in the optimal GNNDR and the corresponding GMI expression is the conditional expectation $\mathbf{E}[\rvx|y, v]$ and the following function:
\begin{eqnarray}
    \omega(y, v) = \mathbf{E}\left[|\rvx|^2 | y, v\right] - \left|\mathbf{E}[\rvx|y, v]\right|^2 = \mathbf{E}\left[\left|\rvx - \mathbf{E}\left[\rvx | y, v\right]\right|^2 \big | y, v\right],
\end{eqnarray}
which {is the variance of $\rvx$ under the conditional probability distribution $p(x|y, v)$, and may also be seen as the conditional mean-squared error (MSE) of the MMSE estimate $\mathbf{E}[\rvx|y, v]$.}

We also examine several restricted and hence suboptimal forms of the GNNDR. The processing and scaling functions, along with the achieved GMIs, can be written in unified closed-form expressions, and are presented in Section \ref{sec:simplied-gnnd}. Here we briefly summarize the various forms of the GNNDR as follows.

\begin{itemize}
    \item Optimal GNNDR: both $g$ and $f$ are general functions of $(y, v)$. The resulting GMI is $I_{\mathrm{GMI, opt}} = \mathbf{E}\left[\log \frac{P}{\omega(\rvy, \rvv)}\right]$. See Theorem \ref{thm:GNND}.
    \item GNNDR with constant scaling function: $g$ is a general function of $(y, v)$ while $f$ is restricted to be a constant. The resulting GMI is $I_{\mathrm{GMI, csf}} = \log \frac{P}{\mathbf{E}[\omega(\rvy, \rvv)]} = \log \frac{P}{\mathsf{mmse}}$ where $\mathsf{mmse}$ is the MMSE of estimating $\rvx$ upon observing $(\rvy, \rvv)$. See Proposition \ref{prop:simplifiya}.
    \item GNNDR with CSI-dependent scaling function: $g$ is a general function of $(y, v)$ while $f$ is restricted to be a general function of $v$ only. The resulting GMI is $I_{\mathrm{GMI, csi}} = \mathbf{E}\left[\log \frac{P}{\mathbf{E}[\omega(\rvy, \rvv)|\rvv]}\right]$. See Proposition \ref{GNNDb}.
    \item GNNDR with linear processing function: $g$ is restricted to be a linear function of $y$ where the linear coefficient vector is a function of $v$, and $f$ is also restricted to be a general function of $v$ only.\footnote{{The considered form of processing and scaling is motivated by conventional approaches in Gaussian fading channels with imperfect CSI, and the obtained $I_{\mathrm{GMI}, \mathrm{lin}}$ revisits some well known results (e.g., \cite{lapidoth02it} \cite[Thm. 2]{weingarten04:it} \cite{hassibi02:it}) in that context; see Section \ref{subsec:case-fading-imperfect-csi}.}} The resulting GMI is $I_{\mathrm{GMI, lin}} = \mathbf{E}\left[\log \frac{P}{\mathsf{lmmse}_\rvv}\right]$ where $\mathsf{lmmse}_\rvv$ is the conditional linear MMSE of estimating $\rvx$ upon observing $(\rvy, \rvv)$, conditioned upon $\rvv$. See Proposition \ref{prop:channel-statec}.
\end{itemize}

Clearly, in general, the following relation holds:
\begin{eqnarray}
    I_{\mathrm{GMI, opt}} \geq I_{\mathrm{GMI, csi}} \geq \max \{ I_{\mathrm{GMI, csf}}, I_{\mathrm{GMI, lin}} \}.
\end{eqnarray}

We also remark that in the absence of channel state, $I_{\mathrm{GMI, csf}}$ and $I_{\mathrm{GMI, lin}}$ degenerate to the results already established in \cite{zhang19jsac}.

In Section \ref{subsec:linear}, we further point out that the {generally suboptimal restricted form of} GNNDR with linear processing function provides a theoretical justification of the commonly adopted, conventional approach of decomposing the channel output $\rvy$ as the sum of $\rvv$-dependent scaled channel input $\rvx$ (``signal part'') and a residual term $\rvw(\rvv)$ (``noise part'') which is conditionally uncorrelated with the signal part conditioned upon $\rvv$. In fact, we show that the resulting GMI, $I_{\mathrm{GMI, lin}}$, coincides with the formally calculated capacity of the decomposed channel when treating the residual term $\rvw(\rvv)$ as Gaussian and independent of the signal part, --- an assumption which is of course not true for general channels. Therefore, the optimal GNNDR improves upon the conventional linear decomposition approach, and suggests how the performance gain may be realized, in principle.

\begin{figure}
    \centering
    \includegraphics[width=4.5in]{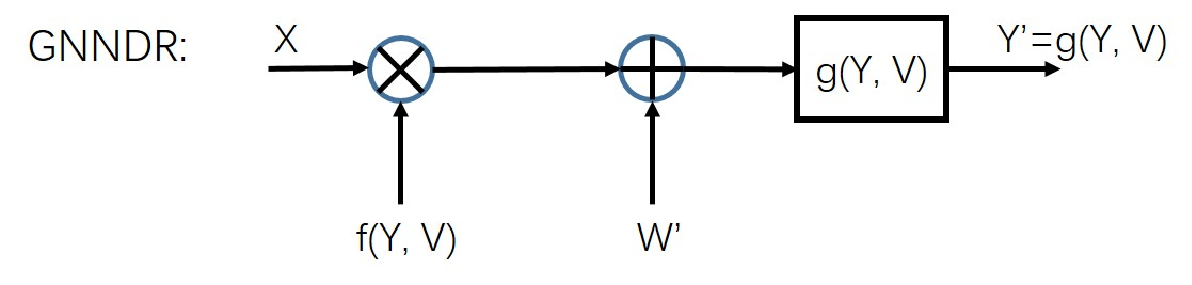}
    \caption{Equivalent channel for GNNDR.}
    \label{fig:GNNDR-decomposition}
    \centering
    \includegraphics[width=3.7in]{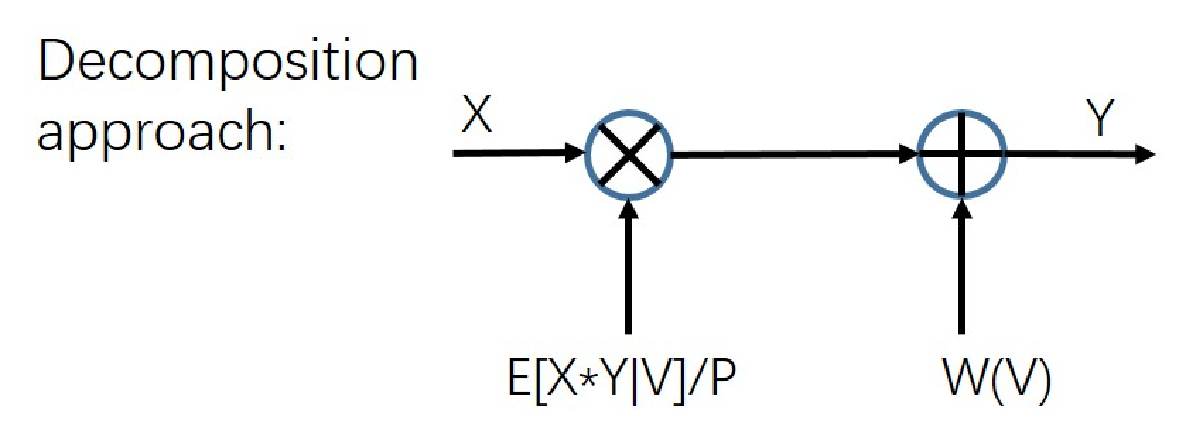}
    \caption{Channel under the {linear} decomposition approach, {which corresponds to the case where $g$ and $f$ are given in Proposition \ref{prop:channel-statec} in Section \ref{subsec:linear}.}}
    \label{fig:Bussgang-decomposition}
\end{figure}

In view of the form of the GNNDR (\ref{eqn:general-nndr}), we may equivalently represent the channel as that illustrated in Figure \ref{fig:GNNDR-decomposition}. The term $\rvw'$ is the difference between the processed channel output, $\rvy' = g(\rvy, \rvv)$, and the scaled channel input, $f(\rvy, \rvv) \rvx$. The GNNDR seeks to minimize the norm of $\rvw'$. In contrast, we illustrate the channel representation under linear decomposition, discussed in the previous paragraph, in Figure \ref{fig:Bussgang-decomposition} (for details see Section \ref{subsec:linear}), where the residual term $\rvw(\rvv)$ is conditionally uncorrelated with the scaled channel input conditioned upon $\rvv$.

\begin{figure}
    \centering
    \includegraphics[width=5.6in]{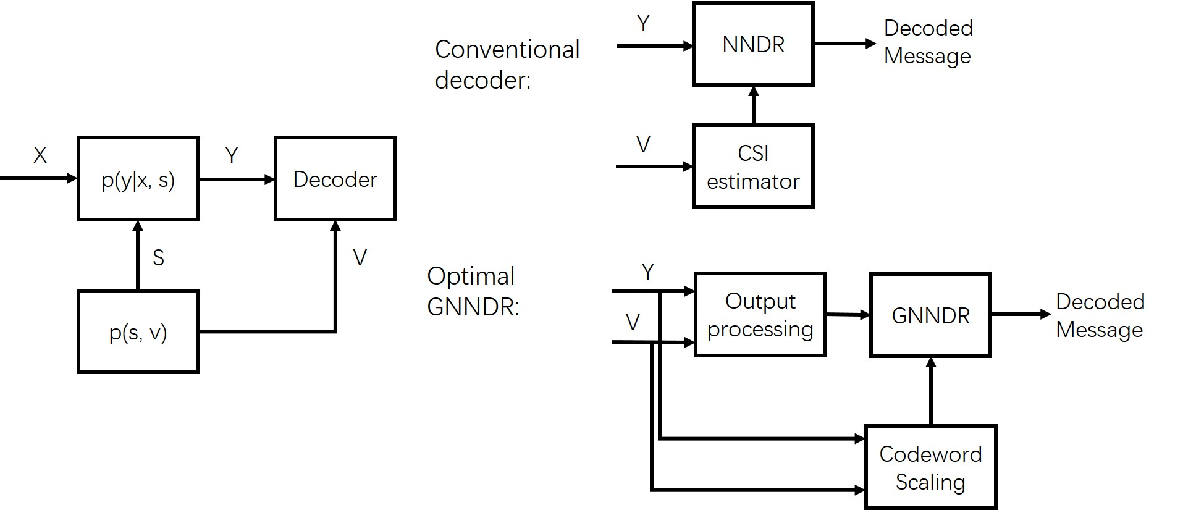}
    \caption{Comparison of the conventional decoder and the optimal GNNDR, for channels with imperfect receiver CSI. The left block diagram is part of the channel model in Figure \ref{fig:model}. The upper right block diagram illustrates the conventional approach for realizing the decoder, where the imperfect receiver CSI is utilized to estimate the true channel state, and the estimated channel state is fed into the NNDR for decoding; the lower right block diagram illustrates the optimal GNNDR, which directly obtains the output processing function and the codeword scaling function, without estimating the channel state. {We remark that channel estimation is still a critical module in practical wireless communication systems, and that the optimal GNNDR structure bypassing the channel estimation step is only valid for the specific information transmission system model in this paper.}}
    \label{fig:imperfect-CSI-decoding}
\end{figure}

{In Section \ref{sec:case-studies}, we illustrate potential applications of the GNNDR via two case studies. The first case study is a Gaussian fading channel with imperfect CSI at the receiver; that is, $\rvs$ and $\rvv$ in Figure \ref{fig:model} are statistically correlated but not necessarily identical. Conventionally, the common approach is to use the imperfect CSI to estimate the true channel state; that is, we form an estimate $\widehat{\rvs}$ based upon $\rvv$, and then treat $\widehat{\rvs}$ as if it is exactly $\rvs$, to return to the scenario where the channel provides perfect CSI to the receiver. For example, when $\rvv$ is a received pilot (polluted by fading and noise), the estimated channel state $\widehat{\rvs}$ is typically the MMSE estimate of $\rvs$ upon observing $\rvv$. From the perspective of the GNNDR, however, it is interesting to notice that the optimal approach to deal with the imperfect CSI $\rvv$ is to directly estimate the channel input $\rvx$, upon observing $(\rvy, \rvv)$, thus skipping the step of estimating the channel state $\rvs$ in the conventional approach. See Figure \ref{fig:imperfect-CSI-decoding} which illustrates the two different approaches. In fact, as we reveal in the case study, the conventional approach is equivalent to restricting the estimtor of $\rvx$ to bear a linear structure, as that in Section \ref{subsec:linear}. Numerical results demonstrate that{even for the familiar Gaussian fading channels without quantization}, adopting the optimal GNNDR yields noticeable performance gain compared with the conventional approach. The second case study is a Gaussian fading channel with one-bit output quantization, with or without dithering. For such a severely distorted nonlinear channel model, the optimal GNNDR again exhibits noticeable performance gain compared with the conventional approach of linear decomposition.}

\subsection{Related Works}
\label{subsec:literature}

Nearest neighbor decoding has been extensively studied in the literature, {and it has inspired numerous decoding algorithms that implement or approximately implement the nearest neighbor search philosophy, e.g., the celebrated Viterbi algorithm for convolutional codes \cite[Chap. 12]{lin04:book}, the ordered statistics decoder (OSD) \cite{fossorier95:it}, the generalized minimum distance (GMD) decoder \cite{forney96:it} \cite{agrawal00:it}, among others.} As we have already pointed out, NNDR is capacity-achieving for Gaussian channels and Gaussian fading channels with perfect receiver CSI. For additive non-Gaussian noise channels, \cite{lapidoth96:it} studied the performance of NNDR under Gaussian input distribution, and established a robustness property that the achieved rate coincides with the channel capacity when the noise distribution is replaced by Gaussian. A second-order asymptotic analysis in this setting was recently conducted in \cite{scarlett17:it}. For Gaussian fading channels with imperfect CSI, treating the imperfectly known fading coefficient as the true one in the NNDR, the robustness of the transceiver structure was studied in \cite{lapidoth02it}, and was extended to multiple-input multiple-output (MIMO) channels by modifying the NNDR to include a weighting matrix \cite{weingarten04:it}. The outage behavior of the NNDR for MIMO Gaussian fading channels with imperfect CSI was studied in \cite{asyhari12:it} \cite{asyhari14:it}. We note that, none of these prior works have considered to generalize the NNDR to incorporate output processing and codeword scaling, as adopted by our approach \cite{zhang16ita} \cite{zhang19jsac} and herein.

For Gaussian fading channels with imperfect CSI, where the imperfect CSI is provided by means like transmitting a prescribed pilot symbol, the related literature is vast; see, e.g., {\cite{medard00:it}} \cite{hassibi02:it} \cite{tong04:spm} {\cite{etkin06:it}} \cite{jindal10:com}. The general design philosophy, which is fundamentally different from and is suboptimal compared with the GNNDR, has been that the channel state, i.e., the fading coefficient in this context, is first estimated from the imperfect CSI, and then treated as if it were the true channel state in the subsequent decoding procedure. We remark that the GNNDR approach may be further combined with the joint processing of pilot and data symbols \cite{valenti01:jsac}-\cite{dorpinghaus12:it}, but we do not pursue this extension in the present paper.

For channels with nonlinear effects, a commonly adopted approach has been decomposing the channel output as the linear superposition of a scaled channel input and a lumped residual term which is uncorrelated with the scaled channel input, and treating the residual term as an additive noise, as we have illustrated in Figure \ref{fig:Bussgang-decomposition}; see, e.g., \cite{ochiai02:com} \cite{bjornson14:it} \cite{orhan15:ita}. Such an approach has its root in \cite{bussgang52} and is thus sometimes termed as ``Bussgang decomposition'' \cite{zhang12com}. As shown in Section \ref{subsec:linear} and briefly discussed in the summary of the GNNDR, this is equivalent to the GNNDR when the output processing function is restricted to be of a linear form.

The GNNDR is a decoding rule with a prescribed structure, and thus its study belongs to the general topic of mismatched decoding \cite{lapidoth98:it} \cite{scarlett20:ftcit}. We briefly introduce the general notion of mismatched decoding in Section \ref{subsec:general-gmi}, which is basically an information transmission model with a prescribed decoding rule declaring the decoded message as the one that minimizes the sum decoding metric, --- generally different from (and hence mismatched to) the log-likelihood function of the channel transition law. In the prior study of mismatched decoding (see, e.g., \cite{lapidoth98:it}-\cite{feldman16:itw} and references therein), the focus has been on deriving various kinds of bounds on the mismatched capacity, by designing different kinds of codebook ensemble. To date, the mismatched capacity remains an open problem in general. In this paper, our work does not contribute to the information theoretic aspect of mismatched decoding, but instead, similar to those in \cite{lapidoth02it} \cite{weingarten04:it} \cite{zhang12com}, applies a known lower bound of the mismatched capacity, namely the GMI, to the GNNDR problem setting. We note that the GMI, which is based upon i.i.d. codebook ensemble, is perhaps the simplest and generally not the best lower bound of the mismatched capacity, but it is applicable to channels with general alphabets and favorable to analysis \cite{ganti00:it}. Furthermore, our work has a key distinction compared to existing works in that we are optimizing among a class of decoding metrics, by maximizing the GMI with respect to the output processing function and the codeword scaling function.

We organize the remaining part of this paper as follows. Section \ref{sec:model} describes the system model. Section \ref{sec:GNND} solves the optimal GNNDR and derives the corresponding GMI. Section \ref{sec:simplied-gnnd} turns to several restricted and suboptimal forms of the GNNDR, establishing their corresponding GMIs. Section \ref{sec:case-studies} presents several case studies, to illustrate possible applications of the obtained forms of the GNNDR. Finally, Section \ref{sec:conclusion} concludes this paper.

\section{System Model}
\label{sec:model}

We consider a general discrete-time memoryless state-dependent channel as already illustrated in Figure \ref{fig:model}, with input $\rvx \in \mathcal{X} = \mathbb{C}$, output $\rvy \in \mathcal{Y}$, state $\rvs \in \mathcal{S}$, and receiver CSI $\rvv \in \mathcal{V}$. The sets $\mathcal{Y}, \mathcal{S}$, and $\mathcal{V}$ are general, not necessarily scalar-valued. When $\rvy$ is arranged into a vector, we use $p$ to denote its dimension.

We assume that the state is independent of the input, and use the channel without feedback. So over a coding block of length $N$, we have
\begin{eqnarray}
P_{\rvs, \rvv}(s^N, v^N) &=& \prod_{n = 1}^N P_{\rvs, \rvv}(s_n, v_n),\\
P_{\rvy|\rvx, \rvs}(y^N|x^N, s^N) &=& \prod_{n = 1}^N P_{\rvy|\rvx, \rvs}(y_n|x_n, s_n).
\end{eqnarray}
Our work can also be extended to the case where the state sequence is stationary ergodic (see, e.g., \cite{lapidoth02it}), whereas for simplicity of exposition we assume that the state sequence is memoryless. Regarding the memoryless assumption, note that we may invoke an ideal interleaver (i.e., with infinite depth and completely random) to create the coding block, so as to render the state sequence memoryless. 

We further assume that at each channel use, $\rvv \leftrightarrow \rvs \leftrightarrow \left(\rvx,\rvy\right)$ is a Markov chain. This implies that the receiver CSI is obtained via some mechanism independent of the current channel use. An example which we study in Section \ref{subsec:case-fading-imperfect-csi} is that the receiver CSI is in the form of received pilot.

We adopt the i.i.d. Gaussian random codebook ensemble. The codebook consists of mutually independent codewords drawn from $\mathcal{CN}(\mathbf{0}, P \mathbf{I}_{N})$ with average power constraint $P$. Given a code rate $R$ (nats/channel use), the encoder uniformly randomly chooses a message $m$ from the message set $\mathcal{M} = \{1, \ldots, \lceil e^{NR}\rceil \}$ for transmission. The encoding function $\mathcal{E}: \mathcal{M} \mapsto \mathcal{X}^N = \mathbb{C}^N$ maps the message $m$ to a length-$N$ codeword $x^N(m)$, which follows $\mathcal{CN}(\mathbf{0}, P \mathbf{I}_{N})$.

Noting that the CSI $\rvv$ is an additional channel output, the decoding function $\mathcal{D}: \mathcal{X}^N \times \mathcal{V}^N \mapsto \mathcal{M}$ maps the channel outputs $(y^N ,v^N)$ to a decoded message $\widehat{m}$. As seen in the introduction, for Gaussian channels with perfect CSI, the optimal, i.e., capacity-achieving, decoder is the NNDR based upon Euclidean distance metrics. In general, despite its possible suboptimality, we adopt the following generalized NNDR (GNNDR):
\begin{eqnarray}
\label{eqn:GNND}
\widehat{m} = \arg \min_{m \in \mathcal{M}} \sum_{n = 1}^N |g(y_n, v_n) - f(y_n, v_n) x_n(m)|^2,
\end{eqnarray}
for a pair of processing function $g$ and scaling function $f$.

In the next section, we use GMI to characterize the achievable rate of the system, and to design the optimal pair of processing and scaling functions.

\section{GMI and Optimal GNNDR}
\label{sec:GNND}

In this section, we introduce the GMI in our setting as a tool to characterize the performance of the system, and solve the problem of maximizing the GMI via optimizing the processing function and the scaling function.

\subsection{GMI of GNNDR}
\label{subsec:general-gmi}

Before studying our problem, we briefly introduce the general notion of mismatched decoding {\cite{ganti00:it} \cite{lapidoth98:it} \cite{scarlett20:ftcit}}. Consider a memoryless channel $P_{\rvy|\rvx}$ with input $\rvx \in \mathcal{X}$ and output $\rvy \in \mathcal{Y}$. At rate $R$ and coding block length $N$, a codebook $\mathcal{C}$ consists of $\lceil e^{NR}\rceil$ codewords, so that the message $m$ is mapped to $x^N(m) = (x_1(m), \ldots, x_N(m))$, for $m \in \mathcal{M} = \{1, \ldots, \lceil e^{NR}\rceil \}$. For mismatched decoding, we let a function $d: \mathcal{X} \times \mathcal{Y} \mapsto \mathbb{R}$ be a so-called ``decoding metric'', which hence induces the following decoding rule:
\begin{eqnarray}
    \label{eqn:sum-metric}
    \mathcal{D}_d: \widehat{m} = \mathrm{arg} \min_{m \in \mathcal{M}} \sum_{n = 1}^N d(x_n(m), y_n),
\end{eqnarray}
with {ties} broken arbitrarily. A rate $R$ is achievable if there exists a sequence of codebooks such that the maximal probability of decoding error asymptotically vanishes as $N \rightarrow \infty$, and the supremum of achievable rates is called the mismatched capacity.

As discussed in Section \ref{subsec:literature}, the mismatched capacity remains an open problem, and various lower bounds to the mismatched capacity have been established, corresponding to different coding schemes. The GMI is such a lower bound, which indicates the achievable rate of mismatched decoding under i.i.d. random codebook ensemble; see, e.g., \cite{ganti00:it} \cite{lapidoth02it}. Indeed, the GMI is the maximum achievable rate such that the probability of decoding error, averaged over the i.i.d. random codebook ensemble, asymptotically vanishes as the coding block length grows without bound \cite{lapidoth02it}.

For a given channel input distribution $P_\rvx$ and its induced channel output distribution $P_\rvy$, a general expression of the GMI is given by \cite{ganti00:it}
\begin{eqnarray}
    \label{eqn:GMI-general-primal}
    I_\mathrm{GMI} = \min_{\nu \in \mathcal{G}} D(\nu \| P_\rvx P_\rvy),
\end{eqnarray}
where $\mathcal{G}$ denotes the set of all probability distributions $\nu$ on $\mathcal{X} \times \mathcal{Y}$ that satisfy
\begin{eqnarray}
    \sum_{x \in \mathcal{X}} \nu(x, y) &=& P_\rvy(y), \quad \forall y \in \mathcal{Y},\\
    \sum_{(x, y) \in \mathcal{X} \times \mathcal{Y}} d(x, y) \nu(x, y) &\leq& \sum_{(x, y) \in \mathcal{X} \times \mathcal{Y}} d(x, y) P_{\rvx, \rvy}(x, y).
\end{eqnarray}
The primal expression (\ref{eqn:GMI-general-primal}) of $I_\mathrm{GMI}$ has an equivalent dual expression as
\begin{eqnarray}
    I_\mathrm{GMI} = \sup_{\theta < 0} \sum_{(x, y) \in \mathcal{X} \times \mathcal{Y}} P_\rvx(x) P_{\rvy|\rvx}(y|x) \log\frac{e^{\theta d(x, y)}}{\sum_{x' \in \mathcal{X}}P_\rvx(x') e^{\theta d(x', y)} }.
\end{eqnarray}

As described in Section \ref{sec:model}, in our context, the channel input follows a circularly symmetric complex Gaussian distribution, and the mismatched decoding rule is the GNNDR (\ref{eqn:GNND}). {Consequently, we have the following general expression of the GMI of the GNNDR.}

\begin{prop}
\label{prop:general GMI}
For the information transmission system model presented in Section \ref{sec:model}, {consider a fixed pair of $g$ and $f$; that is, a decoding metric given by}
{\begin{eqnarray}
    d(x, (y, v)) = |g(y, v) - f(y, v) x|^2.
\end{eqnarray}}
The resulting GMI is given by
\begin{eqnarray}
\label{eqn:GNND-GMI-general}
&&I_{\mathrm{GMI}, g, f}  = \max_{\theta < 0} \bigg\{\theta \mathbf{E}\left[\left|g(\rvy, \rvv) - f(\rvy, \rvv) \rvx \right|^2\right] \nonumber\\
&& - \mathbf{E}\left[\frac{\theta |g(\rvy, \rvv)|^2}{1 - \theta |f(\rvy, \rvv)|^2 P}\right]+ \mathbf{E}\left[\log\left(1 - \theta |f(\rvy, \rvv)|^2 P\right)\right]\bigg\},
\end{eqnarray}
where the expectations are with respect to the joint probability distribution of $\rvv \leftrightarrow \rvs \leftrightarrow (\rvx, \rvy)$, in general.
\end{prop}

{\it Proof:} {We follow similar steps in \cite[Sec. IV]{weingarten04:it}, which illustrates how a random coding argument leads to the expression of the GMI. Consider the evaluation of $P(\widehat{m} \neq m)$, the average decoding error probability over messages and the codebook ensemble. Due to the symmetry in the i.i.d. codebook ensemble, it loses no generality to assume that the codeword corresponding to message $m = 1$ is transmitted (see, e.g., \cite{cover06:book}), i.e.,
\begin{eqnarray}
    P(\widehat{m} \neq m) = P(\widehat{m} \neq 1 | m = 1).
\end{eqnarray}}

{Under $m = 1$, the normalized sum decoding metric in (\ref{eqn:sum-metric}) satisfies
\begin{eqnarray}
    \label{eqn:D1-limit}
    \rvd(1) &=& \frac{1}{N} \sum_{n = 1}^N d(\rvx_n(1), (\rvy_n, \rvv_n))\nonumber\\
    &=& \frac{1}{N} \sum_{n = 1}^N |g(\rvy_n, \rvv_n) - f(\rvy_n, \rvv_n) \rvx_n(1)|^2\nonumber\\
    &\rightarrow& \mathbf{E}\left[|g(\rvy, \rvv) - f(\rvy, \rvv) \rvx|^2\right],
\end{eqnarray}
with probability one, due to the law of large numbers.}

{For an arbitrary $\delta > 0$, define event $\mathcal{A}_\delta = \left\{\rvd(1) \geq \mathbf{E}\left[|g(\rvy, \rvv) - f(\rvy, \rvv) \rvx|^2\right] + \delta\right\}$. The average decoding error probability can be written as
\begin{eqnarray}
    \label{eqn:total-error}
    P(\widehat{m} \neq 1 | m = 1) &=& P(\widehat{m} \neq 1 | m = 1, \mathcal{A}_\delta) P(\mathcal{A}_\delta) + P(\widehat{m} \neq 1 | m = 1, \mathcal{A}^c_\delta) P(\mathcal{A}^c_\delta)\nonumber\\
    &\leq& P(\mathcal{A}_\delta) + P(\widehat{m} \neq 1 | m = 1, \mathcal{A}^c_\delta) P(\mathcal{A}^c_\delta),
\end{eqnarray}
whose first term can be made arbitrarily close to zero for all sufficiently large $N$, according to (\ref{eqn:D1-limit}). Regarding the second term, due to the decoding rule (\ref{eqn:sum-metric}) and the union bound (see, e.g., \cite{cover06:book}), we have
\begin{eqnarray}
    \label{eqn:type-2-error}
    P(\widehat{m} \neq 1 | m = 1, \mathcal{A}^c_\delta) P(\mathcal{A}^c_\delta) &\leq& P\left(\exists m' \neq 1, \rvd(m') < \mathbf{E}\left[|g(\rvy, \rvv) - f(\rvy, \rvv) \rvx|^2\right] + \delta | \mathcal{A}^c_\delta \right) P(\mathcal{A}^c_\delta) \nonumber\\
    &\leq& e^{NR} P\left(\rvd(2) < \mathbf{E}\left[|g(\rvy, \rvv) - f(\rvy, \rvv) \rvx|^2\right] + \delta | \mathcal{A}^c_\delta\right) P(\mathcal{A}^c_\delta) \nonumber\\
    &=& e^{NR} P\left(\rvd(2) < \mathbf{E}\left[|g(\rvy, \rvv) - f(\rvy, \rvv) \rvx|^2\right] + \delta, \mathcal{A}^c_\delta\right) \nonumber\\
    &\leq& e^{NR} P\left(\rvd(2) < \mathbf{E}\left[|g(\rvy, \rvv) - f(\rvy, \rvv) \rvx|^2\right] + \delta\right).
\end{eqnarray}}

{Applying the law of total expectation, we consider
\begin{eqnarray}
    \label{eqn:D2-tail-prob}
    &&P\left(\rvd(2) < \mathbf{E}\left[|g(\rvy, \rvv) - f(\rvy, \rvv) \rvx|^2\right] + \delta\right) \nonumber\\
    &=& \mathbf{E}\left[P\left(\rvd(2) < \mathbf{E}\left[|g(\rvy, \rvv) - f(\rvy, \rvv) \rvx|^2\right] + \delta\Big | (\rvy^N, \rvv^N)\right)\right],
\end{eqnarray}
and proceed to estimating the conditional probability herein.}

{Conditioned upon $(\rvy^N, \rvv^N)$, the normalized sum decoding metric
\begin{eqnarray}
    \rvd(2) = \frac{1}{N} \sum_{n = 1}^N |g(\rvy_n, \rvv_n) - f(\rvy_n, \rvv_n) \rvx_n(2)|^2
\end{eqnarray}
is the average of $N$ independent noncentral chi-square random variables, induced by $\{\rvx_n(2)\}_{n = 1, \ldots, N}$, respectively. So in order to study the asymptotic behavior of its tail probability, we invoke the large deviations principle, specifically the G\"artner-Ellis theorem (see, e.g., \cite[Chap. 2, Sec. 3]{dembo98:book}). For this, we evaluate the conditional moment generating function of $\rvd(2)$, for any $\theta < 0$,
\begin{eqnarray}
    \Lambda_N(N\theta) &=& \log \mathbf{E}\left[e^{N\theta \rvd(2)} \Big | (\rvy^N, \rvv^N)\right]\nonumber\\
    &=& \sum_{n = 1}^N \log \mathbf{E} \left[e^{\theta |g(\rvy_n, \rvv_n) - f(\rvy_n, \rvv_n) \rvx_n(2)|^2} \Big | (\rvy^N, \rvv^N)\right]\nonumber\\
    &=& \sum_{n = 1}^N \left[\frac{\theta |g(\rvy_n, \rvv_n)|^2}{1 - \theta |f(\rvy_n, \rvv_n)|^2 P} - \log \left(1 - \theta |f(\rvy_n, \rvv_n)|^2 P\right)\right],
\end{eqnarray}
where we have utilized the fact that conditioned upon $(\rvy_n, \rvv_n)$, $|g(\rvy_n, \rvv_n) - f(\rvy_n, \rvv_n) \rvx_n(2)|^2$ is a noncentral chi-square random variable with degrees of freedom $2$. Then taking the limit of $\Lambda_N(N\theta)/N$ as $N \rightarrow \infty$, we have
\begin{eqnarray}
    \frac{1}{N} \Lambda_N(N\theta) &=& \frac{1}{N} \sum_{n = 1}^N \left[\frac{\theta |g(\rvy_n, \rvv_n)|^2}{1 - \theta |f(\rvy_n, \rvv_n)|^2 P} - \log \left(1 - \theta |f(\rvy_n, \rvv_n)|^2 P\right)\right]\nonumber\\
    &\rightarrow& \mathbf{E}\left[\frac{\theta |g(\rvy, \rvv)|^2}{1 - \theta |f(\rvy, \rvv)|^2 P}\right] - \mathbf{E}\left[\log \left(1 - \theta |f(\rvy_n, \rvv_n)|^2 P\right)\right]
\end{eqnarray}
with probability one, due to the law of large numbers. Consequently, with probability one, the conditional probability in (\ref{eqn:D2-tail-prob}) exponentially decays to zero at rate
\begin{eqnarray}
    \theta \left(\mathbf{E}\left[|g(\rvy, \rvv) - f(\rvy, \rvv) \rvx|^2\right] + \delta\right) - \mathbf{E}\left[\frac{\theta |g(\rvy, \rvv)|^2}{1 - \theta |f(\rvy, \rvv)|^2 P}\right] + \mathbf{E}\left[\log \left(1 - \theta |f(\rvy_n, \rvv_n)|^2 P\right)\right],
\end{eqnarray}
and hence so does the unconditional probability $P\left(\rvd(2) < \mathbf{E}\left[|g(\rvy, \rvv) - f(\rvy, \rvv) \rvx|^2\right] + \delta\right)$.}

{In view of (\ref{eqn:total-error}) and (\ref{eqn:type-2-error}), we thus have that for any rate $R$ satisfying $R < I_{\mathrm{GMI}, g, f}$ in (\ref{eqn:GNND-GMI-general}), it is possible to find sufficiently small $\delta > 0$ and sufficiently large $N$, such that the average decoding error probability is arbitrarily close to zero. This establishes Proposition \ref{prop:general GMI}. $\square$}

\subsection{GMI-maximizing GNNDR}
\label{subsec:optimal-gnnd}

In this subsection, we investigate the optimal form of the functions $g$ and $f$ so as to maximize the GMI given in Proposition \ref{prop:general GMI}. For
technical convenience, we rewrite $g$ and $f$ as the products of the square root of a real positive function $Q$ called the weighting factor and two distinct functions $\tilde{g}$ and $\tilde{f}$, respectively, i.e.,
\begin{eqnarray}
g(y, v) &=& \sqrt{Q(y, v)} \times \tilde{g}(y, v),\\
f(y, v) &=& \sqrt{Q(y, v)} \times \tilde{f}(y, v).
\end{eqnarray}

Then the decoding metric of the GNNDR (\ref{eqn:GNND}) becomes
\begin{eqnarray}
D(m) &=& \sum_{n = 1}^N |g(y_n, v_n) - f(y_n, v_n) x_n(m)|^2\nonumber\\
&=& \sum_{n = 1}^N Q(y_n, v_n) |\tilde{g}(y_n, v_n) - \tilde{f}(y_n, v_n) x_n(m)|^2,
\end{eqnarray}
and the GMI optimization problem is
\begin{eqnarray}
    &&\max_{g, f} I_{\mathrm{GMI}, g, f} = \max_{\theta < 0, Q > 0, \tilde{g}, \tilde{f}}\nonumber\\
    &&\bigg\{\theta \mathbf{E} \left[Q(\rvy, \rvv) \left|\tilde{g}(\rvy, \rvv)- \tilde{f}(\rvy, \rvv) \rvx \right|^2\right] - \mathbf{E} \left[\frac{\theta Q(\rvy, \rvv) |\tilde{g}(\rvy, \rvv)|^2}{1 - \theta Q(\rvy, \rvv) |\tilde{f}(\rvy, \rvv)|^2 P}\right] \nonumber\\
    &&\quad\quad\quad + \mathbf{E}\left[\log\left(1 - \theta Q(\rvy, \rvv) |\tilde{f}(\rvy, \rvv)|^2 P\right)\right]\bigg\}.
\end{eqnarray}

Absorbing the parameter $\theta < 0$ into $Q > 0$ to rewrite $\tilde{Q} = \theta Q < 0$, and decomposing the overall expectation to swap the order of maximization and conditional expectation with respect to $(\rvy, \rvv)$, the GMI optimization problem becomes
\begin{eqnarray}
    \label{eqn:gmi-optimization-decomposed}
    &&\max_{g, f} I_{\mathrm{GMI}, g, f} = \mathbf{E}_{(\rvy, \rvv)} \max_{\tilde{Q} < 0, \tilde{g}, \tilde{f}} \bigg\{ \tilde{Q}(\rvy,\rvv) \mathbf{E}\left[\left|\tilde{g}(\rvy, \rvv) - \tilde{f}(\rvy, \rvv) \rvx \right|^2 \bigg| \rvy, \rvv \right] \nonumber\\
    &&\quad - \frac{\tilde{Q}(\rvy, \rvv) |\tilde{g}(\rvy, \rvv)|^2}{1 - \tilde{Q}(\rvy, \rvv) |\tilde{f}(\rvy, \rvv)|^2 P} + \log(1 - \tilde{Q}(\rvy, \rvv)|\tilde{f}(\rvy, \rvv)|^2 P)\bigg\}.
\end{eqnarray}
Since $\tilde{Q}, \tilde{g}, \tilde{f}$ are all functions over $\mathcal{Y} \times \mathcal{V}$, we can optimize them for each individual pair of $(y, v)$, and finally take the expectation with respect to $(\rvy, \rvv)$. The resulting optimal solution and the corresponding optimal GNNDR and GMI are given by the following theorem.

\begin{thm}
\label{thm:GNND}
For the information transmission system model presented in Section \ref{sec:model}, the GNNDR that maximizes the GMI in Proposition \ref{prop:general GMI} is given by
\begin{eqnarray}\label{eqn:optimal-generalized-NNDR}
\widehat{m} = \arg\min_{m \in \mathcal{M}} \sum_{n = 1}^{N} \frac{1}{\left(P - \omega(y_n, v_n)\right) \omega(y_n, v_n)} \left|\mathbf{E}[\rvx|y_n, v_n] - \frac{P - \omega(y_n, v_n)}{P} x_n(m)\right|^2,
\end{eqnarray}
and the correspondingly maximized GMI is
\begin{eqnarray}
\label{eqn:max}
I_{\mathrm{GMI, opt}} = \mathbf{E}\left[\log \frac{P}{\omega(\rvy, \rvv)}\right].
\end{eqnarray}
Here, we define
\begin{eqnarray}
\omega(y, v) = \mathbf{E} \left[ |\rvx|^2 | y, v \right] - \left|\mathbf{E} \left[\rvx | y, v \right]\right|^2,
\end{eqnarray}
{and assume that $0 < \omega(\rvy, \rvv) < P$ holds with probability one.}
\end{thm}

{\it Proof:} As said, we optimize the expression (\ref{eqn:gmi-optimization-decomposed}) for each pair of $(y, v)$. For this, let us define it as
\begin{eqnarray}
&&J(\tilde{Q}, \tilde{g}, \tilde{f}) = \tilde{Q}(y, v) \mathbf{E}\left[\left|\tilde{g}(y, v) - \tilde{f}(y, v)\rvx \right|^2 \bigg| y, v\right] \nonumber\\
&&\quad - \frac{\tilde{Q}(y, v) |\tilde{g}(y, v)|^2}{1 - \tilde{Q}(y, v) |\tilde{f}(y, v)|^2 P} + \log\left(1 - \tilde{Q}(y, v) |\tilde{f}(y, v)|^2 P\right).
\end{eqnarray}

Some algebraic manipulations yield
\begin{eqnarray}
&&J(\tilde{Q}, \tilde{g}, \tilde{f}) = \tilde{Q}(y, v) |\tilde{g}(y, v)|^2 + \tilde{Q}(y, v) |\tilde{f}(y, v)|^2 \mathbf{E} \left[ |\rvx|^2 \big|y, v \right] \nonumber\\
&& - 2\tilde{Q}(y, v) \Re\left\{ \tilde{g}^\ast(y, v) \tilde{f}(y, v)\mathbf{E} \left[\rvx \big| y, v \right]\right\} - \frac{\tilde{Q}(y, v) |\tilde{g}(y, v)|^2}{1 - \tilde{Q}(y, v) |\tilde{f}(y, v)|^2 P} + \log\left(1 - \tilde{Q}(y, v) |\tilde{f}(y, v)|^2 P\right)\nonumber\\
&& = \tilde{Q}(y, v) |\tilde{f}(y, v)|^2 \mathbf{E} \left[ |\rvx|^2 \big| y, v \right] - \frac{\tilde{Q}^2(y, v) |\tilde{f}(y, v)|^2 P}{1 - \tilde{Q}(y, v) |\tilde{f}(y, v)|^2 P} |\tilde{g}(y, v)|^2 + \log\left(1 - \tilde{Q}(y, v) |\tilde{f}(y, v)|^2 P\right)\nonumber\\
&& - 2 \tilde{Q}(y, v) \left|\tilde{g}(y, v) \tilde{f}(y, v) \mathbf{E} \left[\rvx \big| y, v \right]\right| \Re\left\{ e^{\jmath \left(\phi\left(\tilde{f}(y, v)\right) + \phi\left(\mathbf{E} \left[\rvx |y, v \right]\right)- \phi\left(\tilde{g}(y, v)\right) \right)} \right\},
\end{eqnarray}
where $\phi$ denotes the phase of its operand complex number. By making a change of variable
\begin{eqnarray}
    \gamma = - \tilde{Q}(y, v) |\tilde{f}(y, v)|^2 P > 0,
\end{eqnarray}
we further rewrite $J(\tilde{Q}, \tilde{g}, \tilde{f})$ as
\begin{eqnarray}
J(\tilde{Q}, \tilde{g}, \phi(\tilde{f}), \gamma) &=& -\frac{\gamma}{P} \mathbf{E} \left[ |\rvx|^2 \big|y, v \right] + \frac{\tilde{Q}(y, v)\gamma}{1 + \gamma} |\tilde{g}(y, v)|^2 + \log (1 + \gamma)\nonumber\\
&& \hspace{-12mm}+ 2\sqrt{\frac{-\gamma \tilde{Q}(y, v)}{P}} \left|\tilde{g}(y, v) \mathbf{E} \left[\rvx \big|y, v \right]\right| \Re\left\{ e^{\jmath \left(\phi\left(\tilde{f}(y, v)\right) + \phi\left(\mathbf{E} \left[\rvx |y, v \right]\right)- \phi\left(\tilde{g}(y, v)\right) \right)} \right\}.
\end{eqnarray}

Letting the partial derivative $\frac{\partial{J}}{\partial{\tilde{Q}}}$ be zero, we find that the optimal $\tilde{Q} < 0$ should satisfy
\begin{eqnarray}\label{eqn:optimal-tilde-Q}
\sqrt{- \tilde{Q}(y, v)} = \frac{(1 + \gamma)\left|\tilde{g}(y, v) \mathbf{E} \left[\rvx \big|y, v \right]\right| \Re\left\{ e^{\jmath \left(\phi\left(\tilde{f}(y, v)\right) + \phi\left(\mathbf{E} \left[\rvx |y, v \right]\right)-\phi\left(\tilde{g}(y, v)\right) \right)} \right\}}{|\tilde{g}(y, v)|^2\sqrt{\gamma P}}.
\end{eqnarray}

Substituting (\ref{eqn:optimal-tilde-Q}) into $J(\tilde{Q}, \tilde{g}, \phi(\tilde{f}), \gamma)$, followed by some algebraic manipulations, we obtain
\begin{eqnarray}
\max_{\tilde{Q}} J(\tilde{Q}, \tilde{g}, \phi(\tilde{f}), \gamma) &=& \log(1 +\gamma) - \frac{\gamma}{P} \mathbf{E} \left[ |\rvx|^2 \big|y, v \right]\nonumber\\
&& + \frac{(1 + \gamma)\bigg[\left|\mathbf{E} \left[\rvx \big|y, v \right]\right| \Re\left\{ e^{\jmath \left(\phi\left(\tilde{f}(y, v)\right) + \phi\left(\mathbf{E} \left[\rvx |y, v \right]\right) - \phi\left(\tilde{g}(y, v)\right) \right)} \right\}\bigg]^2}{P}.
\end{eqnarray}

It is clear that $\max_{\tilde{Q}} J(\tilde{Q}, \tilde{g}, \phi(\tilde{f}), \gamma)$ is further maximized by choosing
\begin{eqnarray}\label{eqn:optimal-phase}
\phi(\tilde{f}(y, v)) = \phi(\tilde{g}(y, v))- \phi(\mathbf{E} \left[\rvx | y, v \right]),
\end{eqnarray}
leading to
\begin{eqnarray}\label{eqn:optimal-phi-f}
\max_{\tilde{Q}, \phi(\tilde{f})} J(\tilde{Q}, \tilde{g}, \phi(\tilde{f}), \gamma) = \log(1 +\gamma) - \frac{\gamma}{P} \mathbf{E} \left[ |\rvx|^2 \big|y, v \right] + \frac{(1 + \gamma) \left|\mathbf{E} \left[\rvx \big| y, v \right]\right|^2}{P},
\end{eqnarray}
which is further independent of $\tilde{g}$.

By maximizing (\ref{eqn:optimal-phi-f}) over $\gamma > 0$, we can find that the optimal $\gamma$ is
\begin{eqnarray}\label{eqn:optimal-gamma}
    \gamma = \frac{P}{\mathbf{E} \left[ |\rvx|^2 | y, v \right] - \left|\mathbf{E} \left[\rvx | y, v \right]\right|^2} - 1 = \frac{P}{\omega(y, v)} - 1.
\end{eqnarray}
Therefore the expression (\ref{eqn:gmi-optimization-decomposed}) maximized for each $(y, v)$ pair is given by
\begin{eqnarray}\label{eqn:optimal-gmi-component}
J(\tilde{Q}, \tilde{g}, \tilde{f}) &=& \log\frac{P}{\omega(y, v)} + \frac{\left|\mathbf{E} \left[ \rvx | y, v \right]\right|^2}{\omega(y, v)} - \frac{P - \omega(y, v)}{P \omega(y, v)} \mathbf{E} \left[ |\rvx|^2 | y, v \right]\nonumber\\
&=& \log\frac{P}{\omega(y, v)} + \frac{\mathbf{E} \left[|\rvx|^2 |y, v \right]}{P} - 1.
\end{eqnarray}

According to (\ref{eqn:gmi-optimization-decomposed}), we then have
\begin{eqnarray}
    \max_{g, f} I_{\mathrm{GMI}, g, f} = \mathbf{E}_{(\rvy, \rvv)} \max_{\tilde{Q}, \tilde{g}, \tilde{f}} J(\tilde{Q}, \tilde{g}, \tilde{f})(\rvy, \rvv),
\end{eqnarray}
{where we have interpreted $J(\tilde{Q}, \tilde{g}, \tilde{f})$ as an operator over $\mathcal{Y} \times \mathcal{V}$.} So applying (\ref{eqn:optimal-gmi-component}), we get
\begin{eqnarray}
    \max_{g, f} I_{\mathrm{GMI}, g, f} &=& \mathbf{E}_{(\rvy, \rvv)} \left[\log\frac{P}{\omega(\rvy, \rvv)} + \frac{\mathbf{E} \left[|\rvx|^2 |\rvy, \rvv \right]}{P} - 1\right]\nonumber\\
    &=& \mathbf{E}\left[\log\frac{P}{\omega(\rvy, \rvv)}\right] + \frac{\mathbf{E}\left[\mathbf{E} \left[|\rvx|^2 |\rvy, \rvv \right]\right]}{P} - 1\nonumber\\
    &=& \mathbf{E}\left[\log\frac{P}{\omega(\rvy, \rvv)}\right],
\end{eqnarray}
where the last equality is due to the law of total expectation, $\mathbf{E}\left[\mathbf{E} \left[|\rvx|^2 |\rvy, \rvv \right]\right] = \mathbf{E}\left[|\rvx|^2\right] = P$.

Tracing back the above proof, we put together (\ref{eqn:optimal-gamma}), (\ref{eqn:optimal-phase}), and (\ref{eqn:optimal-tilde-Q}) to obtain
\begin{eqnarray}
    \label{eqn:optimal-tilde-Q-final}
    \tilde{Q}(y, v) = -\frac{P\left|\mathbf{E}\left[\rvx | y, v \right]\right|^2}{\left|\tilde{g}(y, v)\right|^2 \left(P - \omega(y, v)\right) \omega(y, v)}.
\end{eqnarray}
Recalling the change of variable $\gamma = - \tilde{Q}(y, v) |\tilde{f}(y, v)|^2 P > 0$ leads to
\begin{eqnarray}
    \label{eqn:optimal-tilde-f-magnitude-final}
    \left|\tilde{f}(y, v)\right| = \frac{\left(P - \omega(y, v)\right) \left|\tilde{g}(y, v)\right|}{P \left|\mathbf{E} \left[\rvx |y, v \right]\right|}.
\end{eqnarray}
Combined with (\ref{eqn:optimal-phase}), we obtain
\begin{eqnarray}
    \label{eqn:optimal-tilde-f-final}
    \tilde{f}(y, v) = \frac{\left(P - \omega(y, v)\right) \tilde{g}(y, v)}{P \mathbf{E} \left[\rvx |y, v \right]}.
\end{eqnarray}

From (\ref{eqn:optimal-tilde-Q-final}) and (\ref{eqn:optimal-tilde-f-final}) we can see that the choice of $\tilde{g}(y, v)$ is in fact immaterial, because we may pick an arbitrary $\theta < 0$, say $\theta = -1$ so that $Q(y, v) = -\tilde{Q}(y, v)$, and $\tilde{g}(y, v)$ will disappear when we calculate $g(y, v)$ and $f(y, v)$ via $g(y, v) = \sqrt{Q(y, v)} \times \tilde{g}(y, v)$ and $f(y, v) = \sqrt{Q(y, v)} \times \tilde{f}(y, v)$, respectively. This leads to the GNNDR as given by (\ref{eqn:optimal-generalized-NNDR}), and completes the proof of Theorem \ref{thm:GNND}. $\square$

From the GNNDR (\ref{eqn:optimal-generalized-NNDR}) in Theorem \ref{thm:GNND}, the optimal processing function and scaling function are
\begin{eqnarray}
    \label{eqn:optimal-processing-function}
    g(y, v) &=& \frac{1}{\sqrt{\left(P - \omega(y, v)\right)\omega(y, v)}} \mathbf{E}[\rvx|y, v],\\
    \label{eqn:optimal-scaling-function}
    f(y, v) &=& \frac{\sqrt{P - \omega(y, v)}}{P \sqrt{\omega(y, v)}},
\end{eqnarray}
respectively. We notice the key role played by the conditional expectation $\mathbf{E}[\rvx |y, v]$, which is the MMSE estimate of the channel input upon observing the channel output (including the receiver CSI). This also generalizes the known results reviewed in the background discussion in Section \ref{subsec:motivation}, with the pivotal difference that both $g(y, v)$ and $f(y, v)$ also depend upon the channel output $(y, v)$, via the function $\omega(y, v)$.

Inspecting $\omega(y, v)$, we notice that
\begin{eqnarray}\label{eqn:omega-var-mmse}
    \omega(y, v) = \mathbf{E}\left[|\rvx|^2 | y, v\right] - \left|\mathbf{E}[\rvx|y, v]\right|^2 = \mathbf{E}\left[\left|\rvx - \mathbf{E}[\rvx|y, v]\right|^2 \big| y, v\right],
\end{eqnarray}
{which is the variance of $\rvx$ under the conditional probability distribution $p(x|y, v)$.} Thus by using the law of total expectation, we have
\begin{eqnarray}
    \mathbf{E}\left[\omega(\rvy, \rvv)\right] = \mathbf{E} \left[\left|\rvx - \mathbf{E}[\rvx | \rvy, \rvv]\right|^2\right] = \mathsf{mmse};
\end{eqnarray}
that is, the MMSE of the conditional expectation $\mathbf{E}[\rvx | \rvy, \rvv]$.


\section{{GNNDR with Restricted Forms}}
\label{sec:simplied-gnnd}

{In this section, we turn to several restricted forms of the GNNDR. These restricted forms generally lead to suboptimal performance compared with the optimal GNNDR in Theorem \ref{thm:GNND}, but they may incur less computational cost, and provide further insights into the understanding of the GNNDR.}

\subsection{Constant Scaling Function}
\label{subsec:csf}

In this case, the GNNDR is of the following form:
\begin{eqnarray}
\label{fixed scaling}
\widehat{m}=\arg\min_{m \in \mathcal{M}} \sum_{n = 1}^{N} |g(y_n, v_n) - \alpha x_n(m)|^2,
\end{eqnarray}
where $\alpha \in \mathbb{C}$ is a prescribed constant; that is, here we set the scaling function to be a constant, i.e., $f(y, v) = \alpha$, which should be chosen, together with $g(y, v)$, to maximize the GMI. {Recognizing this case as the model treated in \cite[Prop. 3]{zhang19jsac}, with complex-valued input $\rvx$ and extended output $(\rvy, \rvv)$, we immediately obtain the following result as a corollary of \cite[Prop. 3]{zhang19jsac}.}
\begin{prop}
\label{prop:simplifiya}
For the information transmission system model presented in Section \ref{sec:model}, under decoding rule (\ref{fixed scaling}), the resulting maximized GMI is given by
\begin{eqnarray}
\label{GMIa}
I_{\mathrm{GMI, csf}} = \log \frac{P}{\mathbf{E}\left[\omega(\rvy, \rvv)\right]} = \log \frac{P}{\mathsf{mmse}},
\end{eqnarray}
achieved by
\begin{eqnarray}
\label{eqn:gmi-generic}
g(y, v) &=& \mathbf{E}[\rvx | y, v],\\
\alpha &=& \frac{\mathbf{E}\left[\left|\mathbf{E}[\rvx | \rvy, \rvv]\right|^2\right]}{P} = \frac{P - \mathsf{mmse}}{P}.
\end{eqnarray}
\end{prop}

We may also formally rewrite the processing function and the scaling function {in Proposition \ref{prop:simplifiya} as}
\begin{eqnarray}
    \label{eqn:csf-processing-function}
    g(y, v) &=& \frac{1}{\sqrt{\left(P - \mathbf{E}\left[\omega(\rvy, \rvv)\right]\right) \mathbf{E}\left[\omega(\rvy, \rvv)\right]}} \mathbf{E}[\rvx|y, v],\\
    \label{eqn:csf-scaling-function}
    f(y, v) &=& \frac{\sqrt{P - \mathbf{E}\left[\omega(\rvy, \rvv)\right]}}{P \sqrt{\mathbf{E}\left[\omega(\rvy, \rvv)\right]}},
\end{eqnarray}
so as to compare them with those of the optimal GNNDR (\ref{eqn:optimal-processing-function}) and (\ref{eqn:optimal-scaling-function}) in Theorem \ref{thm:GNND}; that is, here we simply replace all occurrences of $\omega(y, v)$ in (\ref{eqn:optimal-processing-function}) and (\ref{eqn:optimal-scaling-function}) with its expectation, $\mathbf{E}\left[\omega(\rvy, \rvv)\right] = \mathsf{mmse}$.

\subsection{CSI-dependent Scaling Function}
\label{subsec:SI-based}

In this case, we improve the constant scaling function in Section \ref{subsec:csf} by allowing it to depend upon the CSI $\rvv$, but without depending upon the channel output $\rvy$. The GNNDR is of the following form:
\begin{eqnarray}
\label{b}
\widehat{m} = \arg\min_{m \in \mathcal{M}} \sum_{n = 1}^{N} |g(y_n, v_n) - f(v_n) x_n(m)|^2.
\end{eqnarray}
We have the following result regarding the GMI.

\begin{prop}
\label{GNNDb}
For the information transmission system model presented in Section \ref{sec:model}, under decoding rule (\ref{b}), the resulting maximized GMI is given by
\begin{eqnarray}
\label{eqn:2}
I_{\mathrm{GMI, csi}} = \mathbf{E} \left[\log \frac{P}{\mathbf{E}\left[\omega(\rvy, \rvv)|\rvv\right]}\right],
\end{eqnarray}
which is achieved by $g(y, v) = \sqrt{Q(v)} \times \tilde{g}(y, v)$ and $f(v) = \sqrt{Q(v)} \times \tilde{f}(v)$, where
\begin{eqnarray}
\tilde{g}(y, v) &=& \mathbf{E}[\rvx|y, v],\\
\label{f'}
\tilde{f}(v) &=& \frac{\mathbf{E}\left[\rvx^\ast \tilde{g}(\rvy, v) | v \right]}{P},\\
\label{Q}
Q(v) &=& \frac{1}{P \mathbf{E}\left[|\tilde{g}(\rvy, v)|^2 | v \right] - \left|\mathbf{E}\left[\rvx^\ast \tilde{g}(\rvy, v) | v \right]\right|^2}.
\end{eqnarray}
\end{prop}

\textit{Proof:} Similar to the treatment in Section \ref{subsec:optimal-gnnd}, we decompose the processing function and the scaling function into the products of the square root of $Q(v)$ and $\tilde{g}(y, v)$ and $\tilde{f}(v)$, respectively, and then swap the order of maximization and conditional expectation with respect to $\rvv$ when maximizing the GMI. Following steps similar to those in the proof of Theorem \ref{thm:GNND}, for any fixed $\tilde{g}(y, v)$, we can find that the GMI is maximized by choosing
\begin{eqnarray}
Q(v) &=& \frac{1}{P \mathbf{E}\left[|\tilde{g}(\rvy, v)|^2 | v \right] - \left|\mathbf{E}\left[\rvx^\ast \tilde{g}(\rvy, v) | v \right]\right|^2},\\
\tilde{f}(v) &=& \frac{\mathbf{E}\left[\rvx^\ast \tilde{g}(\rvy, v) | v \right]}{P},
\end{eqnarray}
and the corresponding GMI is
\begin{eqnarray}
\label{bb}
I_{\mathrm{GMI}, \tilde{g}} = \mathbf{E}_\rvv\left[\log \frac{1}{1 - \Delta_{\tilde{g}}(\rvv)}\right], \quad \mbox{where}\;  \Delta_{\tilde{g}}(v) = \frac{\left|\mathbf{E}[\rvx^\ast \tilde{g}(\rvy, v)| v]\right|^2}{P \mathbf{E}\left[\left|\tilde{g}(\rvy, v)\right|^2| v\right]}.
\end{eqnarray}

Applying the Cauchy-Schwartz inequality to $\Delta_{\tilde{g}}(v)$, we have
\begin{eqnarray}
\Delta_{\tilde{g}}(v) = \frac{\left|\mathbf{E}\left[\rvx^\ast \tilde{g}(\rvy,v) | v \right]\right|^2}{P \mathbf{E}\left[\left|\tilde{g}(\rvy, v)\right|^2 | v\right]} \leq \frac{\mathbf{E}\left[\left|\mathbf{E}[\rvx | \rvy, v]\right|^2 | v \right]}{P},
\end{eqnarray}
where the equality holds if we let $\tilde{g}(y, v) = \mathbf{E}[\rvx | y, v]$. Hence the maximized GMI becomes
\begin{eqnarray}
    I_{\mathrm{GMI}} = \mathbf{E}\left[\log \frac{P}{P - \mathbf{E}\left[\left|\mathbf{E}[\rvx | \rvy, \rvv]\right|^2 | \rvv \right]}\right].
\end{eqnarray}
The proof of Proposition \ref{GNNDb} is completed by noting that
\begin{eqnarray}\label{eqn:omega-V}
    \mathbf{E}\left[\omega(\rvy, \rvv)|\rvv\right] &=& \mathbf{E}\left[ \mathbf{E} \left[ |\rvx|^2 | \rvy, \rvv \right] - \left|\mathbf{E} \left[\rvx | \rvy, \rvv \right]\right|^2 \bigg| \rvv \right]\nonumber\\
    &=& \mathbf{E} \left[ |\rvx|^2 | \rvv \right] - \mathbf{E}\left[\left|\mathbf{E}[\rvx | \rvy, \rvv]\right|^2 | \rvv \right]\nonumber\\
    &=& P - \mathbf{E}\left[\left|\mathbf{E}[\rvx | \rvy, \rvv]\right|^2 | \rvv \right],
\end{eqnarray}
where we have used the law of total expectation and the fact that $\rvx$ is independent of $\rvv$. $\square$

A further inspection of the functions $\tilde{g}$, $\tilde{f}$ and $Q$ in Proposition \ref{GNNDb} reveals the following relationship:
\begin{eqnarray}
    &&\mathbf{E}\left[\left|\tilde{g}(\rvy, v) - \tilde{f}(v)\rvx \right|^2 \bigg | v \right]\nonumber\\
    &=& \mathbf{E}\left[\left|\tilde{g}(\rvy, v)\right|^2 | v \right] + P |\tilde{f}(v)|^2 - \mathbf{E}\left[\tilde{g}^\ast(\rvy, v) \rvx | v \right] \tilde{f}(v) - \mathbf{E}\left[\tilde{g}(\rvy, v) \rvx^\ast | v \right] \tilde{f}^\ast(v)\nonumber\\
    &=& \mathbf{E}\left[\left|\tilde{g}(\rvy, v)\right|^2 | v \right] + P |\tilde{f}(v)|^2 - P |\tilde{f}(v)|^2 - P |\tilde{f}(v)|^2\nonumber\\
    &=& \mathbf{E}\left[\left|\tilde{g}(\rvy, v)\right|^2 | v \right] - \frac{\left|\mathbf{E}\left[\rvx^\ast \tilde{g}(\rvy, v) | v \right]\right|^2}{P} = \frac{1}{P Q(v)}.
\end{eqnarray}
In other words,
\begin{eqnarray}
\label{eqn:qq}
Q(v) = \frac{1/P}{\mathbf{E}\left[\left|\tilde{g}(\rvy, v) - \tilde{f}(v)\rvx\right|^2 \bigg | v \right]}.
\end{eqnarray}
The denominator, $\mathbf{E}\left[\left|\tilde{g}(\rvy, v) - \tilde{f}(v)\rvx\right|^2 \bigg | v \right]$, tracks the mean-squared difference between $\tilde{g}(\rvy, v) = \mathbf{E}\left[\rvx | \rvy, v\right]$ and $\tilde{f}(v) \rvx$. Hence the effect of $Q(v)$ is essentially a ``normalizer'' for each value of $v$, the receiver CSI.

Besides, we recognize from (\ref{eqn:omega-V}) that $\mathbf{E}\left[\omega(\rvy, \rvv)|\rvv\right]$ is in fact equal to the conditional MMSE,
\begin{eqnarray}
    \mathbf{E}\left[\left|\rvx - \mathbf{E}\left[\rvx |\rvy, \rvv\right]\right|^2 \bigg| \rvv\right] = \mathsf{mmse}_\rvv,
\end{eqnarray}
and hence we can rewrite the GMI in Proposition \ref{GNNDb} as
\begin{eqnarray}
    \label{eqn:gmi-csi-mmse}
    I_{\mathrm{GMI}, \mathrm{csi}} = \mathbf{E}\left[\log \frac{P}{\mathsf{mmse}_\rvv}\right].
\end{eqnarray}

\subsection{Linear Processing Function}
\label{subsec:linear}

In this subsection, we simplify the processing function by restricting it to be a linear function of $y$ upon observing $v$; that is, $g(y, v) = \beta^\ast(v) y$, where $\beta(\cdot)$ is a column vector function of the CSI $v$ and we use the inner product between $\beta(v)$ and $y$ as the processing function. We further restrict the scaling function to be a function of $v$ only, as that in Section \ref{subsec:SI-based}.

Again, we decompose $g$ and $f$ into $g(y, v) = \sqrt{Q(v)} \times \tilde{\beta}^\ast(v) y$ and $f(v) = \sqrt{Q(v)} \times \tilde{f}(v)$. The resulting GMI is given by the following proposition.

\begin{prop}
\label{prop:channel-statec}
For the information transmission system model presented in Section \ref{sec:model}, under linear processing function, the resulting maximized GMI is given by
\begin{eqnarray}
\label{eqn:GMI-linear}
I_{\mathrm{GMI}, \mathrm{lin}} = \mathbf{E} \left[\log \frac{P}{P - \mathbf{E}[\rvx^\ast \rvy|\rvv]^\ast \mathbf{E}[\rvy\rvy^\ast | \rvv]^{-1} \mathbf{E}[\rvx^\ast \rvy|\rvv]}\right],
\end{eqnarray}
achieved by $g(y, v) = \sqrt{Q(v)} \times \tilde{\beta}^\ast(v) y$ and $f(v) = \sqrt{Q(v)} \times \tilde{f}(v)$, where
\begin{eqnarray}
\tilde{\beta}(v) &=& \mathbf{E}[\rvy\rvy^\ast|v]^{-1} \mathbf{E}[\rvx^\ast \rvy|v],\\
\tilde{f}(v) &=& \frac{\mathbf{E}[\rvx^\ast \rvy| v]^\ast \mathbf{E}[\rvy\rvy^\ast | v]^{-1} \mathbf{E}[\rvx^\ast \rvy| v]}{P},\\
\label{eqn:Q-lin}
Q(v) &=& \frac{1}{\mathbf{E}[\rvx^\ast \rvy| v]^\ast \mathbf{E}[\rvy\rvy^\ast | v]^{-1} \mathbf{E}[\rvx^\ast \rvy| v] \left(P - \mathbf{E}[\rvx^\ast \rvy| v]^\ast \mathbf{E}[\rvy\rvy^\ast | v]^{-1} \mathbf{E}[\rvx^\ast \rvy| v]\right)}.
\end{eqnarray}
Here we have assumed that for the information transmission system, $\mathbf{E}[\rvy\rvy^\ast | v]$ is invertible for any $v \in \mathcal{V}$.
\end{prop}

\textit{Proof:} We resume from (\ref{bb}) in the proof of Proposition \ref{GNNDb}. By substituting the linear processing function $\tilde{g}(y, v) = \tilde{\beta}(v)^\ast y$ into $Q(v)$, $\tilde{f}(v)$, and $\Delta_{\tilde{g}}(v)$, we have,
\begin{eqnarray}
    Q(v) &=& \frac{1}{P \mathbf{E}\left[|\tilde{g}(\rvy, v)|^2 | v \right] - \left|\mathbf{E}\left[\rvx^\ast \tilde{g}(\rvy, v) | v \right]\right|^2}\nonumber\\
    &=& \frac{1}{\tilde{\beta}^\ast(v) \left(P \mathbf{E}\left[\rvy \rvy^\ast | v \right] - \mathbf{E}\left[\rvx^\ast \rvy | v\right] \mathbf{E}\left[\rvx^\ast \rvy | v\right]^\ast\right)\tilde{\beta}(v)},\\
    \tilde{f}(v) &=& \frac{\mathbf{E}\left[\rvx^\ast \tilde{g}(\rvy, v) | v \right]}{P} = \frac{\tilde{\beta}^\ast(v) \mathbf{E}\left[\rvx^\ast \rvy | v \right]}{P},\\
    \label{eqn:Delta-generalized-Rayleigh-quotient}
    \Delta_{\tilde{g}}(v) &=& \frac{\left|\mathbf{E}[\rvx^\ast \tilde{g}(\rvy, v)| v]\right|^2}{P \mathbf{E}\left[\left|\tilde{g}(\rvy, v)\right|^2| v\right]}
    = \frac{\tilde{\beta}^\ast(v) \mathbf{E}\left[\rvx^\ast\rvy|v\right]\mathbf{E}\left[\rvx^\ast\rvy|v\right]^\ast \tilde{\beta}(v)}{P \tilde{\beta}^\ast(v) \mathbf{E}\left[\rvy\rvy^\ast|v\right] \tilde{\beta}(v)}.
\end{eqnarray}

Note that $\mathbf{E}\left[\rvx^\ast \rvy| v\right] \mathbf{E}\left[\rvx^\ast \rvy | v\right]^\ast$ and $\mathbf{E}\left[\rvy\rvy^\ast| v\right]$ are both Hermitian matrices. By recognizing (\ref{eqn:Delta-generalized-Rayleigh-quotient}) as a generalized Rayleigh quotient, we follow the same argument as that in the proof of \cite[Prop. 2]{zhang19jsac}, to {transform the generalized Rayleigh quotient into a standard Rayleigh quotient (see \cite[Eqn. (12)]{zhang19jsac}) and} obtain that
\begin{eqnarray}
    \max_{\tilde{g}} \Delta_{\tilde{g}}(v) = \mathbf{E}\left[\rvx^\ast\rvy|v\right]^\ast \mathbf{E}\left[\rvy\rvy^\ast|v\right]^{-1} \mathbf{E}\left[\rvx^\ast\rvy|v\right],
\end{eqnarray}
achieved by $\tilde{\beta}(v) = \mathbf{E}\left[\rvy\rvy^\ast|v\right]^{-1} \mathbf{E}\left[\rvx^\ast\rvy|v\right]$. This completes the proof of Proposition \ref{prop:channel-statec}. $\square$

From standard linear estimation theory \cite{poorbook}, we immediately recognize that the denominator in $I_{\mathrm{GMI}, \mathrm{lin}}$ (\ref{eqn:GMI-linear}), $P - \mathbf{E}[\rvx^\ast \rvy|\rvv]^\ast \mathbf{E}[\rvy\rvy^\ast | \rvv]^{-1} \mathbf{E}[\rvx^\ast \rvy|\rvv]$, is exactly the conditional MMSE of the linear MMSE estimator of $\rvx$ upon observing $\rvy$, conditioned upon $\rvv$. Hence we rewrite the GMI in Proposition \ref{prop:channel-statec} as
\begin{eqnarray}
I_{\mathrm{GMI}, \mathrm{lin}} = \mathbf{E}\left[\log \frac{P}{\mathsf{lmmse}_\rvv}\right].
\end{eqnarray}
Compared with $I_{\mathrm{GMI}, \mathrm{csi}}$ (\ref{eqn:gmi-csi-mmse}) in Section \ref{subsec:SI-based}, we clearly see the performance loss due to replacing the MMSE estimator by the linear MMSE estimator for $\tilde{g}$.

We can also interpret $Q(v)$ in (\ref{eqn:Q-lin}) as a ``normalizer'' akin to that in Section \ref{subsec:SI-based}, by rewriting $Q(v)$ in Proposition \ref{prop:channel-statec} as
\begin{eqnarray}
    Q(v) = \frac{1/P}{\mathbf{E}\left[\left|\tilde{\beta}^\ast(v) \rvy - \tilde{f}(v) \rvx\right|^2  \bigg | v\right]}.
\end{eqnarray}

In the following, we provide a heuristic point of view for Proposition \ref{prop:channel-statec}, related to the so-called Bussgang decomposition. Such a point of view has been described for channels with scalar-valued output and without CSI in \cite{zhang19jsac}, and here we extend it in a general sense. For each given value of $v$, we may follow the idea of the Bussgang decomposition to express $\rvy$ as
\begin{eqnarray}
\rvy = \frac{\mathbf{E}[\rvx^\ast \rvy|v]}{P} \rvx + \rvw(v),\label{eqn:bussgang1}
\end{eqnarray}
where the residual term, i.e., the ``noise part'', vector $\rvw(v) = \rvy - \frac{\mathbf{E}[\rvx^\ast \rvy|v]}{P} \rvx$ can be shown to be conditionally uncorrelated with the ``signal part'' $\frac{\mathbf{E}[\rvx^\ast \rvy|v]}{P} \rvx$, i.e., $\mathbf{E}\left[\frac{\mathbf{E}[\rvx^\ast \rvy|v]}{P} \rvx \rvw^\ast(v) \bigg | v\right] = \mathbf{0}$. Furthermore, we can verify that the mean vector and the covariance matrix of $\rvw(v)$ are zero and $\mathbf{E}\left[\rvy \rvy^\ast \big |v\right] - \frac{1}{P}\mathbf{E}\left[\rvx^\ast \rvy \big | v\right] \mathbf{E}\left[\rvx^\ast \rvy \big | v\right]^\ast$, respectively.

Hence by viewing (\ref{eqn:bussgang1}) as a linearized channel, and applying a whitening filter to $\rvy$, we may formally derive the CSI-dependent ``signal-to-noise ratio'' (SNR) of (\ref{eqn:bussgang1}) as
\begin{eqnarray}
    \label{eqn:Bussgang-SNR}
    \mathsf{snr}(v) &=& \left|\left(\mathbf{E}\left[\rvy \rvy^\ast \big |v\right] - \frac{1}{P}\mathbf{E}\left[\rvx^\ast \rvy \big | v\right] \mathbf{E}\left[\rvx^\ast \rvy \big | v\right]^\ast\right)^{-1/2} \frac{\mathbf{E}[\rvx^\ast \rvy|v]}{P} \right|^2 P\nonumber\\
    &=& \frac{1}{P} \mathbf{E}[\rvx^\ast \rvy|v]^\ast \left(\mathbf{E}\left[\rvy \rvy^\ast \big |v\right] - \frac{1}{P}\mathbf{E}\left[\rvx^\ast \rvy \big | v\right] \mathbf{E}\left[\rvx^\ast \rvy \big | v\right]^\ast\right)^{-1} \mathbf{E}[\rvx^\ast \rvy|v]\nonumber\\
    &=& \frac{\mathbf{E}[\rvx^\ast \rvy|v]^\ast \mathbf{E}[\rvy \rvy^\ast |v]^{-1} \mathbf{E}[\rvx^\ast \rvy|v]}{P - \mathbf{E}[\rvx^\ast \rvy|v]^\ast \mathbf{E}[\rvy \rvy^\ast |v]^{-1} \mathbf{E}[\rvx^\ast \rvy|v]},
\end{eqnarray}
by invoking the Sherman-Morrison formula followed by some algebraic manipulations.

Comparing (\ref{eqn:Bussgang-SNR}) with the GMI expression (\ref{eqn:GMI-linear}) in Proposition \ref{prop:channel-statec}, we immediately have the following relationship hold:
\begin{eqnarray}
I_{\mathrm{GMI}, \mathrm{lin}} = \mathbf{E} \left[\log(1 + \mathsf{snr}(\rvv)\right].
\end{eqnarray}
This thus provides a theoretic justification of the Bussgang decomposition; that is, by lumping the effect of (possibly noisy) nonlinearity as an overall residual noise term uncorrelated with the channel input, we can theoretically guarantee the achievable rate of $I_{\mathrm{GMI}, \mathrm{lin}}$, which, however, is generally lower than $I_{\mathrm{GMI}, \mathrm{csi}}$ and $I_{\mathrm{GMI}, \mathrm{opt}}$.

\section{{Case Studies}}
\label{sec:case-studies}

{In this section, we provide two case studies for illustrating the results developed in the past two sections.}

\subsection{{Fading Channels with Imperfect CSI}}
\label{subsec:case-fading-imperfect-csi}

Consider a Gaussian fading channel
\begin{eqnarray}
\label{eqn:channel}
    \rvy = \rvs \rvx + \rvz,
\end{eqnarray}
where $\rvx \sim \mathcal{CN}(0, P)$ and $\rvy \in \mathbb{C}^p$. The Gaussian noise vector $\rvz \sim \mathcal{CN}(\mathbf{0}, \sigma^2 \mathbf{I}_p)$ and the fading vector $\rvs$ are independent of $\rvx$. The CSI $\rvv$ at the receiver is assumed to be a general random variable correlated with $\rvs$, such that $\rvv \leftrightarrow \rvs \leftrightarrow (\rvx, \rvy)$, as assumed in Section \ref{sec:model}.

We have the following expressions of the GMI.
\begin{prop}
    \label{prop:GMI-fading-imperfect-csi}
    For the Gaussian fading channel model (\ref{eqn:channel}), we have
    \begin{eqnarray}
    \label{eqn:opt}
    I_{\mathrm{GMI}, \mathrm{opt}} &=& \mathbf{E}\left[\log \frac{P}
    {\mathbf{E}_{\rvs |\rvy,\rvv}\left[\frac{P\sigma^2}{\sigma^2 + P |\rvs|^2}\right]+
    \mathrm{var}_{\rvs |\rvy,\rvv}\left(\frac{P\rvs^\ast \rvy}{\sigma^2 + P |\rvs|^2} \right)}\right],\\
    \label{V_C_GMI}
    I_\mathrm{GMI, lin} &=& \mathbf{E}\left[\log \left(1 + P\widehat{\rvs}^\ast\left(\sigma^2\mathbf{I} + P\mathbf{E}[(\rvs-\widehat{\rvs}) (\rvs-\widehat{\rvs})^\ast | \widehat{\rvs}]\right)^{-1}\widehat{\rvs} \right)\right],
    \end{eqnarray}
    where $\widehat{\rvs} = \mathbf{E}\left[\rvs | \rvv\right]$ and $\mathbf{E}[ (\rvs-\widehat{\rvs} ) (\rvs-\widehat{\rvs} )^\ast |\widehat{\rvs} ]$ is the error covariance matrix.
\end{prop}

\textit{Proof:} Applying Theorem \ref{thm:GNND}, we have
\begin{eqnarray}
    I_{\mathrm{GMI}, \mathrm{opt}} &=& \mathbf{E}\left[\log \frac{P}{\omega(\rvy, \rvv)}\right] \nonumber\\
    &=& \mathbf{E}\left[\log \frac{P}{\mathbf{E}_{\rvs |\rvy,\rvv}\left[\mathrm{var}\left(\rvx |\rvy,\rvs\right)\right]+
    \mathrm{var}_{\rvs |\rvy,\rvv}\left(\mathbf{E}\left[\rvx |\rvy,\rvs\right]\right)}\right] \nonumber\\
    &=& \mathbf{E}\left[\log \frac{P}
    {\mathbf{E}_{\rvs |\rvy,\rvv}\left[\frac{P\sigma^2}{\sigma^2 + P |\rvs|^2}\right]+
    \mathrm{var}_{\rvs |\rvy,\rvv}\left(\frac{P\rvs^\ast \rvy}{\sigma^2 + P |\rvs|^2} \right)}\right],
\end{eqnarray}
where we have applied the law of total variance and the Sherman-Morrison formula.

On the other hand, regarding the general expression of $I_\mathrm{GMI, lin}$ in Proposition \ref{prop:channel-statec}, we begin with
\begin{eqnarray}
    \label{eqn:sim}
    \mathbf{E}[\rvx^\ast \rvy |\rvv] &=& P\mathbf{E}[\rvs |\rvv], \\
    \mbox{and}\;\;\mathbf{E}[\rvy \rvy^\ast |\rvv] &=& P\mathbf{E}[\rvs \rvs^\ast |\rvv]+ \sigma^2 \mathbf{I},
\end{eqnarray}
whose inverse can be written as
\begin{eqnarray}
    \left(P\mathbf{E}[\rvs \rvs^\ast |\rvv]+ \sigma^2 \mathbf{I}\right)^{-1} &=& (\underbrace{\sigma^2 \mathbf{I} + P\mathbf{E} [(\rvs-\widehat{\rvs})(\rvs-\widehat{\rvs})^\ast |\widehat{\rvs}]}_{\rva} + P\widehat{\rvs}  \widehat{\rvs}^\ast)^{-1}\nonumber\\
    &=& \rva^{-1} - \frac{P \rva^{-1} \widehat{\rvs} \widehat{\rvs}^\ast \rva^{-1}} {1 + P\widehat{\rvs}^\ast \rva^{-1} \widehat{\rvs}}.
\end{eqnarray}
So with some algebraic manipulations, it follows that
\begin{eqnarray}
    I_\mathrm{GMI, lin} &=& \mathbf{E}\left[\log \left(\frac{P}{P-\mathbf{E}\left[\rvx^\ast\rvy|\rvv\right]^\ast\mathbf{E}\left[\rvy\rvy^\ast|\rvv\right]^{-1}\mathbf{E}\left[\rvx^\ast\rvy|\rvv\right]} \right)\right] \nonumber\\
    &=& \mathbf{E}\left[\log \frac{1}{1 - P\widehat{\rvs}^\ast \rva^{-1}\widehat{\rvs} + P^2 \widehat{\rvs}^\ast
    \frac{\rva^{-1} \widehat{\rvs} \widehat{\rvs}^\ast \rva^{-1}}{1 + P\widehat{\rvs}^\ast \rva^{-1} \widehat{\rvs}}\widehat{\rvs}} \right]\nonumber\\
    &=& \mathbf{E}\left[\log \left(1 + P\widehat{\rvs}^\ast\left(\sigma^2\mathbf{I} + P\mathbf{E}[(\rvs-\widehat{\rvs}) (\rvs-\widehat{\rvs})^\ast | \widehat{\rvs}]\right)^{-1}\widehat{\rvs} \right)\right].
\end{eqnarray}
This completes the proof of Proposition \ref{prop:GMI-fading-imperfect-csi}. $\square$

For $I_\mathrm{GMI, opt}$, we note that although there is no nonlinear effect in the channel model, the optimal GNNDR is nonlinear in $(y, v)$, because $(\rvx, \rvy, \rvv)$ are generally not jointly Gaussian, for example, as generated by (\ref{eqn:channel}) and (\ref{eqn:pilot}) below.

For $I_\mathrm{GMI, lin}$, according to the matrix determinant lemma, we may rewrite (\ref{V_C_GMI}) as
\begin{eqnarray}
    I_\mathrm{GMI, lin} = \mathbf{E}\left[\log \det\left( \mathbf{I}+P\left(\sigma^2 \mathbf{I} + P\mathbf{E} [(\rvs - \widehat{\rvs}) (\rvs-\widehat{\rvs})^\ast |\widehat{\rvs} ]\right)^{-1} \widehat{\rvs}\widehat{\rvs}^\ast \right)\right],
\end{eqnarray}
which is exactly the well known achievable rate derived in \cite[Thm. 2]{weingarten04:it} (see also \cite{lapidoth02it}) via a GMI analysis, and \cite{hassibi02:it} via a worst-case noise argument, when specialized to the scalar-input channel model (\ref{eqn:channel}), by the following linear decomposition:
\begin{eqnarray}
\label{eqn:decomposition-error-as-residual-noise}
\rvy = \rvs \rvx + \rvz = \widehat{\rvs} \rvx + (\tilde{\rvs} \rvx + \rvz).
\end{eqnarray}

If $\rvs$ consists of independent components, the error covariance matrix $\mathbf{E}[ (\rvs-\widehat{\rvs} ) (\rvs-\widehat{\rvs} )^\ast |\widehat{\rvs} ]$ is diagonal. Furthermore, assume that $\rvv$ is provided to the receiver as a received pilot,\footnote{Here, in order to be consistent with the general system model in Section \ref{sec:model}, we do not include the transmission of the pilot symbol in the channel uses. If a pilot symbol is inserted every $\tau$ channel uses, then the resulting achieved information rate should be discounted by a factor of $(1 - 1/\tau)$, of course.}
\begin{eqnarray}
    \label{eqn:pilot}
    \rvv = \rvs_\mathrm{p} x_\mathrm{p} + \rvz_\mathrm{p},
\end{eqnarray}
where $x_\mathrm{p}$ is the prescribed transmitted pilot symbol, and $\rvs_\mathrm{p}$ and $\rvz_\mathrm{p}$ are the fading vector and the noise vector when receiving the pilot, respectively. Under Rayleigh fading where $\rvs$ and $\rvs_\mathrm{p}$ are $\mathcal{CN}(\mathbf{0}, \eta^2 \mathbf{I})$ and correlated, $\rvs$ and $\rvv$ are jointly Gaussian. Consequently, $\widehat{\rvs}$ is the linear MMSE estimate of $\rvs$, for which the error covariance matrix becomes
\begin{eqnarray}
    \mathbf{E}[ (\rvs-\widehat{\rvs} ) (\rvs-\widehat{\rvs} )^\ast |\widehat{\rvs} ] = \eta^2 \mathbf{I} - \frac{|x_\mathrm{p}|^2 \mathbf{E}\left[\rvs \rvs_\mathrm{p}^\ast\right] \mathbf{E}\left[\rvs_\mathrm{p} \rvs^\ast\right]}{|x_\mathrm{p}|^2\eta^2 + \sigma^2}.
\end{eqnarray}

\begin{figure}
    \centering
    \includegraphics[width=5in]{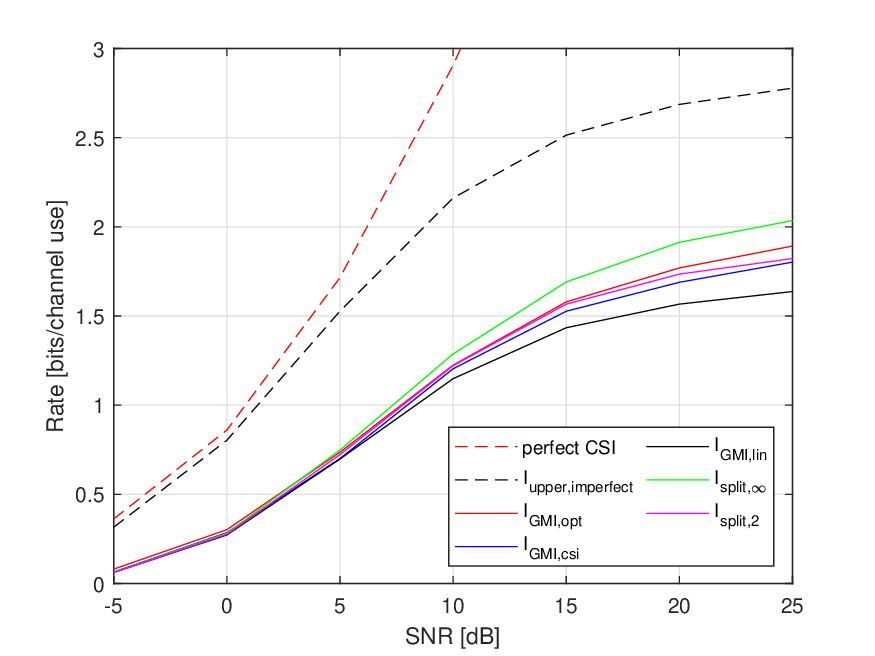}
    \caption{GMIs for Gaussian fading channel (\ref{eqn:channel}) with imperfect CSI (\ref{eqn:pilot}), $x_\mathrm{p} = \sqrt{P/2} + \jmath \sqrt{P/2}$, $\mathbf{E}\left[\rvs_\mathrm{p} \rvs^\ast\right] = 0.8584$.}
    \label{fig:linear_ergodic_high}
\end{figure}

\begin{figure}
    \centering
    \includegraphics[width=5in]{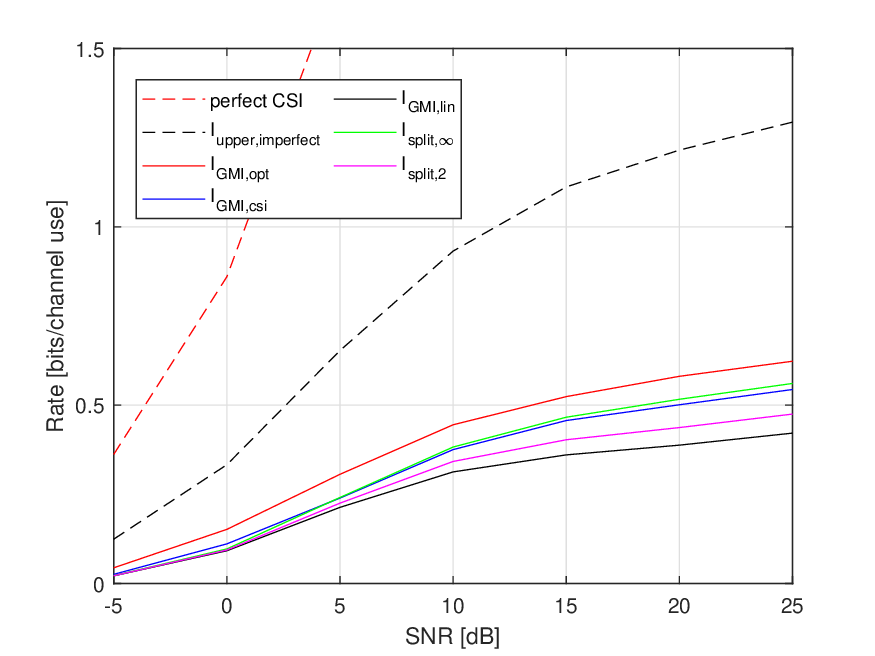}
    \caption{GMIs for Gaussian fading channel (\ref{eqn:channel}) with imperfect CSI (\ref{eqn:pilot}), $x_\mathrm{p} = \sqrt{P/2} + \jmath \sqrt{P/2}$, $\mathbf{E}\left[\rvs_\mathrm{p} \rvs^\ast\right] = 0.5046$.}
    \label{fig:linear_ergodic_low}
\end{figure}

Figures \ref{fig:linear_ergodic_high} and \ref{fig:linear_ergodic_low} compare the achieved GMIs for the optimal GNNDR (Theorem \ref{thm:GNND} and Proposition \ref{prop:GMI-fading-imperfect-csi}), the GNNDR with CSI-dependent scaling function (Proposition \ref{GNNDb}), and the GNNDR with linear processing function (Propositions \ref{prop:channel-statec} and \ref{prop:GMI-fading-imperfect-csi}), for single-antenna receiver (i.e., $p = 1$) with fading correlation coefficients $\mathbf{E}\left[\rvs_\mathrm{p} \rvs^\ast\right] = 0.8584$ and $0.5046$, respectively. We observe that $I_{\mathrm{GMI}, \mathrm{opt}}$ evidently outperforms $I_{\mathrm{GMI}, \mathrm{lin}}$, and that $I_{\mathrm{GMI}, \mathrm{csi}}$ lies between them. Their gaps increase as the fading correlation coefficient decreases, suggesting that the benefit of the optimal GNNDR is more essential as the channel becomes less predictable.

For comparison we also plot the capacity of (\ref{eqn:channel}) with perfect receiver CSI, and an upper bound on the channel mutual information with imperfect receiver CSI under Gaussian input \cite[Lem. 6.2.1 and Sec. VI-C]{lapidoth02it}:
\begin{eqnarray}
    I(\rvx; \rvy, \rvv) \leq \log\left(1+\frac{P}{\sigma^2}\right) - \mathbf{E}\left[\log\left(1+\frac{|\rvx|^2}{\sigma^2}\right)\right] + I(\rvs;\rvv),
\end{eqnarray}
which, for our channel model, can be shown to be bounded in the SNR $P/\sigma^2$;\footnote{In fact, under Gaussian input, the first two terms converge to Euler's constant $\gamma \approx 0.577$, and with $x_\mathrm{p} = \sqrt{P/2} + \jmath \sqrt{P/2}$, the third term has the following limit,
\begin{eqnarray}
    \lim_{P/\sigma^2 \rightarrow \infty} I(\rvs; \rvv) = \log\left(\frac{\eta^2}{\eta^2 - \frac{|\mathbf{E}\left[\rvs_\mathrm{p} \rvs^\ast\right]|^2}{\eta^2}}\right).
\end{eqnarray}} see, e.g., \cite[Sec. VI-C]{lapidoth02it}.

{Another comparison we include in Figures \ref{fig:linear_ergodic_high} and \ref{fig:linear_ergodic_low} is the achievable rates with a rate-splitting technique \cite{pastore14:it}. As mentioned in Section \ref{subsec:literature}, structured codebook ensemble may lead to performance improvement beyond the i.i.d. codebook ensemble used in our GMI analysis, and the rate-splitting technique utilizes layered encoding and decoding to introduce structure.\footnote{{In Figures \ref{fig:linear_ergodic_high} and \ref{fig:linear_ergodic_low}, $I_{\mathrm{split}, 2}$ means that there are two layers and $I_{\mathrm{split}, \infty}$ means that the number of layers tends to infinity.}} Note that the rate-splitting
technique and the GNNDR are not competitive approaches, but can be complementary. Therefore, an interesting future direction is to integrate the GNNDR into the decoding of the rate-splitting technique for further enhancement.}


\subsection{{Channels with One-bit Output Quantization}}
\label{subsec:case-one-bit}

Here, we quantize the channel output of (\ref{eqn:channel}) by an one-bit quantizer. To simplify the situation so as to focus on the impact of quantization, we assume that the CSI is perfect, i.e., $\rvv = \rvs$.

When the one-bit quantizer is symmetric, it produces the quantized output at the $i$-th receive antenna $\rvy_i = \rvy_i^\mathrm{R} + \jmath \rvy_i^\mathrm{I} \in \mathbb{C}$, $i = 1, \ldots, p$, as
\begin{eqnarray}
\label{eqn:1bit-no-state}
    \rvy_i^\mathrm{R} = \mathrm{sgn}(\rvs_i^\mathrm{R} \rvx^\mathrm{R}- \rvs_i^\mathrm{I} \rvx^\mathrm{I} + \rvz_i^\mathrm{R}), \quad \rvy_i^\mathrm{I} = \mathrm{sgn}(\rvs_i^\mathrm{R} \rvx^\mathrm{I} + \rvs_i^\mathrm{I} \rvx^\mathrm{R} + \rvz_i^\mathrm{I}),
\end{eqnarray}
where $\mathrm{sgn}(x) = 1$ if $x \geq 0$ and $-1$ otherwise for $x \in \mathbb{R}$.

We also consider adding a bias to the received signal prior to quantization. This technique is called dithered quantization \cite{gray98it}, which, if appropriately exercised, renders the quantization error to behave in a desirable statistical pattern. {Here we consider deterministic dithering so that the added bias is a prescribed quantity.} The dithered quantized output is
\begin{eqnarray}
\label{eqn:1bit-no-state-dither}
    \rvy_i^\mathrm{R} = \mathrm{sgn}(\rvs_i^\mathrm{R} \rvx^\mathrm{R} - \rvs_i^\mathrm{I} \rvx^\mathrm{I} + \rvz_i^\mathrm{R} + b_i^\mathrm{R}), \quad \rvy_i^\mathrm{I} = \mathrm{sgn}(\rvs_i^\mathrm{R} \rvx^\mathrm{I} + \rvs_i^\mathrm{I} \rvx^\mathrm{R} + \rvz_i^\mathrm{I} + b_i^\mathrm{I}), \quad i = 1, \ldots, p.
\end{eqnarray}
Here $b_i^{\mathrm{R}/\mathrm{I}}$ is a prescribed dither added to the $i$-th received antenna. An exhaustive search of optimal dithers is prohibitive, and we instead adopt a heuristic design as $b_i = \alpha \sqrt{P/2} \rvs_i t_i$, where $t_i$ is the solution of $\Psi(t) = i/(p + 1)$, $\Psi(t)$ is the cumulative distribution function of the standard real Gaussian distribution $\mathcal{N}(0, 1)$, and $\alpha$ is a parameter which can be numerically optimized \cite{zhang19jsac}.

\begin{figure}
    \centering
    \includegraphics[width=5in]{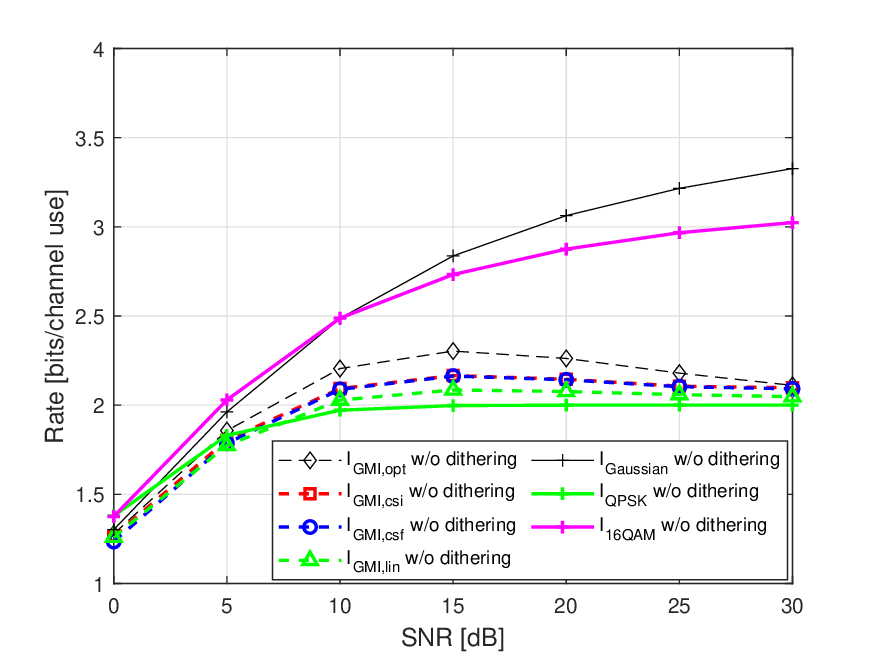}
    \caption{GMI and mutual information for channel (\ref{eqn:1bit-no-state}) without dithering, i.i.d. Rayleigh fading, $p = 4$.}
    \label{fig:InVar_1Q_2D}
\end{figure}

\begin{figure}
    \centering
    \includegraphics[width=5in]{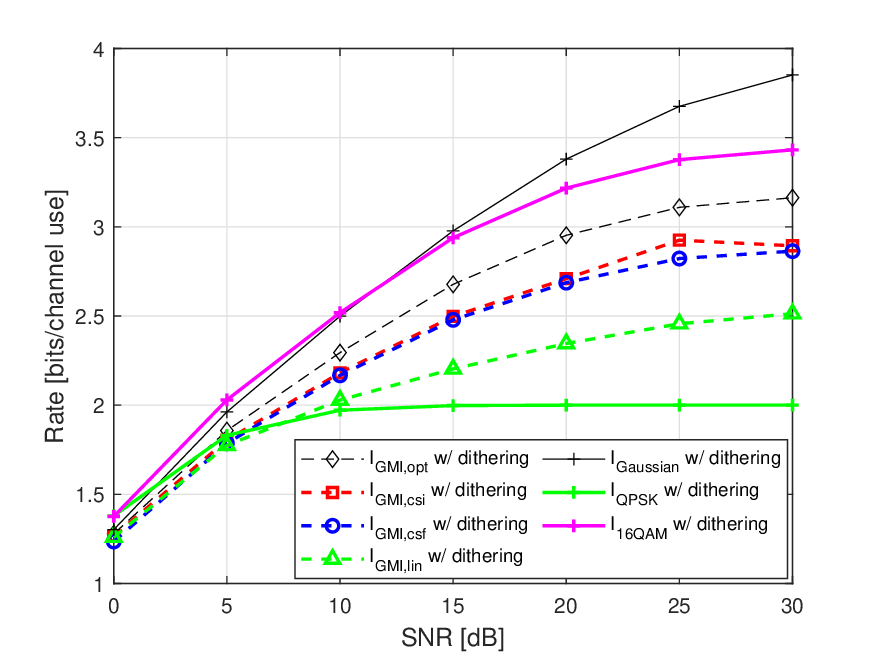}
    \caption{GMI and mutual information for channel (\ref{eqn:1bit-no-state-dither}) with dithering, i.i.d. Rayleigh fading, $p = 4$.}
    \label{fig:InVar_1Q_4D}
\end{figure}

Figures \ref{fig:InVar_1Q_2D} and \ref{fig:InVar_1Q_4D} display GMI and mutual information of the one-bit quantized channel with and without dithering, (\ref{eqn:1bit-no-state}) and (\ref{eqn:1bit-no-state-dither}), respectively. The number of receive antennas is $p = 4$, subject to i.i.d. Rayleigh fading. In each figure, we observe that the gap between $I_{\mathrm{GMI, opt}}$ and $I_{\mathrm{GMI, lin}}$ increases as SNR increases, and that the curves of $I_{\mathrm{GMI, csf}}$ and $I_{\mathrm{GMI, csi}}$ are fairly close, lying in between the curves of $I_{\mathrm{GMI, opt}}$ and $I_{\mathrm{GMI, lin}}$. Furthermore, comparing Figures \ref{fig:InVar_1Q_2D} and \ref{fig:InVar_1Q_4D}, we observe that dithering is an effective technique to boost the GMI.

For comparison, we also plot the mutual information achieved by QPSK, 16QAM, and Gaussian channel inputs. Recall that the GNNDR is a mismatched decoder, while achieving mutual information requires a matched decoder such as the maximum likelihood decoder. We observe that the GNNDR outperforms QPSK, and that the gap between GMI of the GNNDR and the mutual information of 16QAM/Gaussian is substantially reduced by dithering.

\section{Conclusion}
\label{sec:conclusion}

In this paper, we have introduced several generalizations of the celebrated nearest neighbor decoding rule. These generalizations are unified under the framework of the GNNDR. Via output processing and codeword scaling, the GNNDR effectively aligns the channel input to match the transformed channel output, in such a way that the performance measure of GMI is maximized. The GNNDR provides a systematic approach to handling the possibly nonlinear channel effects, and to exploiting the possibly imperfect receiver CSI. Two notable consequences are in order. First, compared with the conventional approach of decomposing the channel output into the linear superposition of a scaled channel input and an uncorrelated residual term, the GNNDR leads to improved performance. Second, compared with the conventional approach where the channel state is first estimated and then treated as if it is perfect in decoding, it is beneficial for the receiver to directly estimate the channel input and perform the GNNDR. These shed new insights into the architecture of transceiver design.

We mention two possible extensions for future study. First, if some form of CSI is also available at the transmitter \cite{caire99:it}, then the problem formulation may be extended to include the possibility of link adaptation. Second, when the transmitter is equipped with multiple antennas and wishes to transmit multiple information streams, we need to work with a {multi-dimensional} version of the GNNDR; see, e.g., \cite{weingarten04:it} {\cite{pang21:itw}}.

In closing, we briefly comment on challenges for implementing the GNNDR. First, our analysis and results hinge on the assumption of Gaussian input distribution, so a natural next step is to solve for the optimal or nearly optimal GNNDR for practical discrete input distributions. Second, for general channels, direct calculation of the conditional expectations in the GNNDR typically involves numerical multiple integrations and thus is not usually tractable. We note, however, that conditional expectation is closely related to regression, and therefore, a promising approach is to apply computational tools from the vast literature on nonlinear regression \cite{hastie:book}, for example, kernel estimators and neural networks; see, e.g., \cite{zhang19jsac}.


\begin{thebibliography}{1}
\bibliographystyle{IEEEtran}

    \bibitem{lapidoth96:it} A.~Lapidoth, ``Nearest-neighbor decoding for additive non-Gaussian noise channels,'' \emph{IEEE Trans. Inform. Theory}, 42(5), 1520-1529, Sep. 1996.

    \bibitem{bjornson14:it} E.~Bj\"ornson, J.~Hoydis, M.~Kountouris, and M.~Debbah, ``Massive MIMO systems with non-ideal hardware: energy efficiency, estimation, and capacity limits,'' \emph{IEEE Trans. Inform. Theory}, 60(11), 7112-7139, Nov. 2014.

    \bibitem{zhang12com} W.~Zhang, ``A general framework for transmission with transceiver distortion and some applications,'' \emph{IEEE Trans. Commun.}, 60(2), 384-399, Feb. 2012.

    \bibitem{zhang16ita} W.~Zhang, ``A remark on channels with transceiver distortion,'' in \emph{Proc. Inf. Theory Appl. Workshop (ITA)}, pp. 1-4, Jan. 2016.

    \bibitem{zhang19jsac} W.~Zhang, Y.~Wang, C.~Shen and N.~Liang, ``A regression approach to certain information transmission problems,'' \emph{IEEE J. Select. Areas Commun.}, 37(11), 2517-2531, Nov. 2019.
    
    \bibitem{lapidoth02it} A.~Lapidoth and S.~Shamai (Shitz), ``Fading channels: how perfect need 'perfect side information' be?'' \emph{IEEE Trans. Inform. Theory}, 48(5), 1118-1134, May 2002.
    
    \bibitem{weingarten04:it} H.~Weingarten, Y.~Steinberg, and S.~Shamai (Shitz), ``Gaussian codes and weighted nearest neighbor decoding in fading multiple antenna channels,'' \emph{IEEE Trans. Inform. Theory}, 50(8), 1665-1686, Aug. 2004.

    \bibitem{hassibi02:it} B.~Hassibi and B.~Hochwald, ``How much training is needed in multiple-antenna wireless links?'' \emph{IEEE Trans. Inform. Theory}, 49(4),  951-963, Apr. 2003.

    \bibitem{lin04:book} {S.~Lin and D.~J.~Costello, \emph{Error Control Coding}, 2nd ed., Upper Saddle River, NJ, USA: Prentice Hall, 2004.}

    \bibitem{fossorier95:it} {M.~P.~C.~Fossorier and S.~Lin, ``Soft-decision decoding of linear block codes based on ordered statistics,'' \emph{IEEE Trans. Inform. Theory}, 41(5), 1379-1396, Sep. 1995.}

    \bibitem{forney96:it} {G.~D.~Forney and A.~Vardy, ``Generalized minimum-distance decoding of Euclidean-space codes and lattices,'' \emph{IEEE Trans. Inform. Theory}, 42(6), 1992-2026, Nov. 1996.}

    \bibitem{agrawal00:it} {D.~Agrawal and A.~Vardy, ``Generalized minimum distance decoding in Euclidean space: performance analysis,'' \emph{IEEE Trans. Inform. Theory}, 46(1), 60-83, Jan. 2000.}

    \bibitem{scarlett17:it} J.~Scarlett, V.~Y.~F.~Tan and G.~Durisi, ``The dispersion of nearest-neighbor decoding for additive non-Gaussian channels,'' \emph{IEEE Trans. Inform. Theory}, 63(1), 81-92, Jan. 2017.

    \bibitem{asyhari12:it} A.~T.~Asyhari and A.~G.~i F\`{a}bregas, ``Nearest neighbor decoding in MIMO block-fading channels with imperfect CSIR,'' \emph{IEEE Trans. Inform. Theory}, 58(3), 1483-1517, Mar. 2012.
    
    \bibitem{asyhari14:it} A.~T.~Asyhari and A.~G.~i F\`{a}bregas, ``MIMO block-fading channels with mismatched CSI,'' \emph{IEEE Trans. Inform. Theory}, 60(11), 7166-7185, Nov. 2014.

    \bibitem{medard00:it} {M.~M\'{e}dard, ``The effect upon channel capacity in wireless communications of perfect and imperfect knowledge of the channel,'' \emph{IEEE Trans. Inform. Theory}, 46(3), 933-946, May 2000.}

    \bibitem{tong04:spm} L.~Tong, B.~M.~Sadler, and M.~Dong, ``Pilot-assisted wireless transmissions: general model, design criteria, and signal processing,'' \emph{IEEE Signal Process. Magazine}, 21(6), 12-25, Nov. 2004.

    \bibitem{etkin06:it} {R.~H.~Etkin and D.~N.~C.~Tse, ``Degrees of freedom in some underspread MIMO fading channels,'' \emph{IEEE Trans. Inform. Theory}, 52(4), 1576-1608, Apr. 2006.}

    \bibitem{jindal10:com} N.~Jindal and A.~Lozano, ``A unified treatment of optimum pilot overhead in multipath fading channels,'' \emph{IEEE Trans. Commun.}, 58(10), 2939-2948, Oct. 2010.

    \bibitem{valenti01:jsac} M.~C.~Valenti and B.~D.~Woerner, ``Iterative channel estimation and decoding of pilot symbol assisted turbo codes over flat-fading channels,'' \emph{IEEE J. Sel. Areas Commun.}, 19(9), 1697-1705, Sep. 2001.

    \bibitem{zhang07:it} W.~Zhang and J.~N.~Laneman, ``How good is PSK for peak-limited fading channels in the low-SNR regime?'' \emph{IEEE Trans. Inform. Theory}, 53(1), 236-251, Jan. 2007.

    \bibitem{jindal09:isit} N.~Jindal, A.~Lozano, and T.~Marzetta, ``What is the value of joint processing of pilots and data in block-fading channels?'' in \emph{Proc. IEEE Int. Symp. Inf. Theory (ISIT)}, Seoul, Korea, Jun. 2009.
    
    \bibitem{dorpinghaus12:it} M.~Dorpinghaus, A.~Ispas, and H.~Meyr, ``On the gain of joint processing of pilot and data symbols in stationary Rayleigh fading channels,'' \emph{IEEE Trans. Inform. Theory}, 58(5), 2963-2982, May 2012.

    \bibitem{ochiai02:com} H.~Ochiai and H.~Imai, ``Performance analysis of deliberately clipped OFDM signals,'' \emph{IEEE Trans. Commun.}, 50(1), 89-101, Jan. 2002.

    \bibitem{orhan15:ita} O.~Orhan, E.~Erkip, and S.~Rangan, ``Low power analog-to-digital conversion in millimeter wave systems: impact of resolution and bandwidth on performance,'' in \emph{Proc. Inf. Theory Appl. Workshop (ITA)}, 191-198, Feb. 2015.

    \bibitem{bussgang52} J.~J.~Bussgang, ``Crosscorrelation functions of amplitude-distorted Gaussian signals,'' Massachusetts Inst. Technol., Cambridge, MA, USA, Tech. Rep. TR 216, Mar. 1952.

    \bibitem{lapidoth98:it} A.~Lapidoth and P.~Narayan, ``Reliable communication under channel uncertainty,'' \emph{IEEE Trans. Inform. Theory}, 44(6), 2148-2177, Oct. 1998.

    \bibitem{scarlett20:ftcit} J.~Scarlett, A.~G.~i F\`{a}bregas, A.~Somekh-Baruch, and A.~Martinez, ``Information-theoretic foundations of mismatched decoding,'' \emph{Foundations and Trends in Communications and Information Theory}, 17(2-3), 149-401, 2020.

    \bibitem{stiglitz65:it} I.~G.~Stiglitz, ``Coding for a class of unknown channels,'' \emph{IEEE Trans. Inform. Theory}, 12(2), 189-195, Apr. 1966.

    \bibitem{csiszar81:it} I.~Csisz\'{a}r and J.~K\"{o}rner, ``Graph decomposition: a new key to coding theorems,'' \emph{IEEE Trans. Inform. Theory}, 27(1), 5-12, Jan. 1981.

    \bibitem{hui:phd} J.~Y.~N.~Hui, ``Fundamental issues of multiple accessing,'' Ph.D. dissertation, Mass. Inst. Technol., Cambridge, MA, USA, 1983.

    \bibitem{merhav94:it} N.~Merhav, G.~Kaplan, A.~Lapidoth, and S.~Shamai (Shitz), ``On information rates for mismatched decoders,'' \emph{IEEE Trans. Inform. Theory}, 40(6), 1953-1967, Nov. 1994.
    
    \bibitem{csiszar95:it} I.~Csisz\'{a}r and P.~Narayan, ``Channel capacity for a given decoding metric,'' \emph{IEEE Trans. Inform. Theory}, 41(1), 35-43, Jan. 1995.

    \bibitem{salz95:com} J.~Salz and E.~Zehavi, ``Decoding under integer metrics constraints,'' \emph{IEEE Trans. Commun.}, 43(2), 307-317, Feb. 1995.

    \bibitem{ganti00:it} A.~Ganti, A.~Lapidoth, and \'{I}.~E.~Telatar, ``Mismatched decoding revisited: general alphabets, channels with memory, and the wide-band limit,'' \emph{IEEE Trans. Inform. Theory}, 46(7), 2315-2328, Nov. 2000.

    \bibitem{scarlett:phd} J.~Scarlett, ``Reliable communication under mismatched decoding,'' Ph.D. dissertation, Univ. Cambridge, Cambridge, U.K., 2014.

    \bibitem{pastore14:it} {A.~Pastore, T.~Koch, and J.~R.~Fonollosa, ``A rate-splitting approach to fading channels with imperfect channel-state information,'' \emph{IEEE Trans. Inform. Theory}, 60(7), 2014.}

    \bibitem{somekh15:it} {A.~Somekh-Baruch, ``A general formula for the mismatched capacity,'' \emph{IEEE Trans. Inform. Theory}, 61(9), 4554-4568, Sep. 2015.}

    \bibitem{feldman16:itw} {Y.~Feldman and A.~Somekh-Baruch, ``Channels with state information and mismatched decoding,'' \emph{IEEE Inform. Theory Workshop (ITW)}, 419-423, 2016.}

    \bibitem{cover06:book} T.~M.~Cover and J.~A.~Thomas, \textit{Elements of Information Theory}, 2nd ed., Hoboken, NJ, USA: John Wiley \& Sons, 2006.

    \bibitem{dembo98:book} A.~Dembo and O.~Zeitouni, \textit{Large Deviations Techniques and Applications}, 2nd ed., Berlin, Heidelberg: Springer, 1998.

    \bibitem{poorbook} H.~V.~Poor, \textit{An Introduction to Signal Detection and Estimation}, 2nd ed., New York, NY, USA: Springer, 1994.

    \bibitem{gray98it} R.~M.~Gray and D.~L.~Neuhoff, ``Quantization,'' \emph{IEEE Trans. Inform. Theory}, 44(6), 2325-2383, Oct. 1998.

    \bibitem{caire99:it} G.~Caire and S.~Shamai (Shitz), ``On the capacity of some channels with channel state information,'' \emph{IEEE Trans. Inform. Theory}, 45(6), 2007-2019, Sep. 1999.

    \bibitem{pang21:itw} {S.~Pang and W.~Zhang, ``Generalized nearest neighbor decoding for MIMO channels with imperfect channel state information,'' \emph{IEEE Inform. Theory Workshop (ITW)}, 2021.}

    \bibitem{hastie:book} T.~Hastie, R.~Tibshirani, and J.~Friedman, \textit{The Elements of Statistical Learning}, 2nd ed., New York, NY, USA: Springer, 2009.

\end{thebibliography}
\end{document}